\documentclass[12pt, leqno]{article}
\usepackage{amsmath,setspace,multirow,lineno}
\usepackage[font=small,labelfont=bf]{caption}
\usepackage[top=1.1in, bottom=1.1in, left=1.1in, right=1.1in]{geometry}

\usepackage[round]{natbib}
\usepackage{color,soul}

\DeclareCaptionStyle{italic}[justification=centering]{labelfont={bf},textfont={it},labelsep=colon}
\usepackage{graphicx,psfrag,epsf,textcomp,epstopdf, amsthm}
\usepackage{enumerate, dsfont}
\usepackage{natbib}
\usepackage{url,xcolor}
\usepackage{booktabs, subfig, bm, paralist,mathpazo,tikz,todonotes,longtable,microtype}
\usepackage[linesnumbered,ruled,vlined]{algorithm2e}

\usepackage[pdftex,colorlinks=true]{hyperref}
\definecolor{darkblue}{rgb}{0,0,.6}
\hypersetup{citecolor=darkblue,linkcolor=darkblue,urlcolor=darkblue}
\definecolor{DarkRed}{rgb}{.7,0,.4}

\usepackage{comment}

\newcommand{\blind}{0}

\newcommand{\X}{\mathcal{X}}
\newcommand{\Y}{\mathcal{Y}}

\graphicspath{{plots/}}
\DeclareMathOperator*{\argmin}{\arg\!\min}

\newsavebox\CBox

 \newtheorem{@definition}{\sc Definition}[section]

  \renewcommand\X{\mathcal{X}}

\date{}

\begin{document}

\def\spacingset#1{\renewcommand{\baselinestretch}{#1}\small\normalsize} \spacingset{1}

\if0\blind
{
\title{\bf Function-on-function partial quantile regression}}
\author{
Ufuk Beyaztas \footnote{Corresponding address: Department of Statistics, Marmara University, Istanbul, Turkey; Email: ufuk.beyaztas@marmara.edu.tr; ORCID: \url{https://orcid.org/0000-0002-5208-4950}} 
\\
Department of Statistics \\
Marmara University \\
\\
Han Lin Shang \\
    Department of Actuarial Studies and Business Analytics \\
    Macquarie University \\
\\
Aylin Alin \\
    Department of Statistics \\
    Dokuz Eylul University \\
}
\maketitle
\fi

\if1\blind
{
\title{\bf Function-on-function partial quantile regression}
} \fi

\maketitle

\begin{abstract}
A function-on-function linear quantile regression model, where both the response and predictors consist of random curves, is proposed by extending the classical quantile regression setting into the functional data to characterize the entire conditional distribution of functional response. In this paper, a functional partial quantile regression approach, a quantile regression analog of the functional partial least squares regression, is proposed to estimate the function-on-function linear quantile regression model. A partial quantile covariance function is first used to extract the functional partial quantile regression basis functions. The extracted basis functions are then used to obtain the functional partial quantile regression components and estimate the final model. Although the functional random variables belong to an infinite-dimensional space, they are observed in a finite set of discrete-time points in practice. Thus, in our proposal, the functional forms of the discretely observed random variables are first constructed via a finite-dimensional basis function expansion method. The functional partial quantile regression constructed using the functional random variables is approximated via the partial quantile regression constructed using the basis expansion coefficients. The proposed method uses an iterative procedure to extract the partial quantile regression components. A  Bayesian information criterion is used to determine the optimum number of retained components. The proposed functional partial quantile regression model allows for more than one functional predictor in the model. However, the true form of the proposed model is unspecified, as the relevant predictors for the model are unknown in practice. Thus, a forward variable selection procedure is used to determine the significant predictors for the proposed model. Moreover, a case-sampling-based bootstrap procedure is used to construct pointwise prediction intervals for the functional response. The predictive performance of the proposed method is evaluated using several Monte Carlo experiments under different data generation processes and error distributions. The finite-sample performance of the proposed method is compared with the functional partial least squares method. Through an empirical data example, air quality data are analyzed to demonstrate the effectiveness of the proposed method.
\end{abstract}


\spacingset{1.45} 

\section{Introduction} \label{sec:intro}

Conditional mean regression is a general framework to find the tendency and average relationship between the response and predictor/s variables in many life sciences, such as agriculture, epidemiology, hydrology, health, and biology. However, there is also a strong need to explore the effects of predictors on the response at non-central locations of the response variable's distribution in some cases. Most importantly, in many fields, researchers are interested in the extreme events (i.e., the upper and lower tails of the distribution) to understand better the effects of predictors on the entire conditional distribution of the response variable. For example, from an agricultural policy perspective, one may be interested in the upper tail of the conditional farmland price distribution to determine the most important price drivers of the more expensive farmland \citep[see, e.g.,][]{Lehn}. In a climate-related study, the climatic variables recorded at climatically different regions (for example, extremely rainy and dry regions) need to be analyzed at different regimes. For example, climatic variables belonging to the rainy regions need to be analyzed at the upper tail of the distribution while variables obtained from dry regions need to be analyzed at the lower tail of the distribution. Climatically different regions require different risk management \citep[see, e.g.,][]{Abbas}. In a study investigating the effects of air pollution on birth weight, one may be interested in the change in different levels of the birth weight distribution associated with a change in exposure to air pollutants \citep[see, e.g.,][]{Lara}. Alternatively, in an air-pollution-related study, the primary interest may be to predicting higher concentration levels of the pollutants \citep[see, e.g.,][]{Vasseur}. The examples can be extended to other scientific fields \citep[see, e.g.,][]{Eilers, Briollais, Magzamen}.

The conditional mean regression is limited in the above cases because the relationship between the response and predictors estimated from the mean regression can not be easily extended to non-central locations of the response variable's distribution. As an alternative to conditional mean regression, quantile regression (QR) proposed by \cite{Koenker1978} evaluates the impacts of predictors on the response at different quantile levels. Besides its ability to characterize the entire conditional distribution of the response variable, QR has several other desirable properties not provided by mean regression. For instance:
\begin{inparaenum}
\item[1)] QR belongs to a robust family \citep{Koenker2005}, and thus, it provides a more robust inference than mean regression in the presence of outliers.
\item[2)] Unlike mean regression, QR does not assume any particular distribution for the response variable and error terms. Consequently, compared with mean regression, more efficient results may be obtained using QR when the error term follows a non-Gaussian heavy-tailed distribution.
\item[3)] Contrary to mean regression, QR is a non-parametric approach, and it does not require the constant variance assumption for the response variable. Therefore, it provides a more efficient inference than mean regression in the presence of heteroskedasticity.
\end{inparaenum}
Consult \cite{Koenker2005} for more information about the theoretical properties and empirical applications of QR.

The traditional regression models discussed above are optimum for analyzing the discretely observed data. On the other hand, technological developments in data collection tools in the last few decades have led to an increase in functional data whose elements are recorded over a continuum in many fields of science. Existing traditional methods are not capable of analyzing such data due to the infinite-dimensional nature of the functional variables. Therefore, the need for developing statistical analysis techniques to analyze functional data has been increased \citep[see, e.g.,][]{ferraty2005, ramsay2006, cuevas2014, horvath2012}. Among many others, the functional linear regression models (FLRMs), in which at least one of the response or predictor variables involve infinite-dimensional random curves, have been extensively used to explore the association between the response and predictor variables \citep[see e.g.,][and references therein for recent studies about the FLRMs]{chiou2016, Bande2017, greven2017, cao2018, BSenv, BeyaztasShang2021}.

QR has led to interesting studies in the FLRMs. For the QR settings in scalar-on-function regression models, where the response is scalar and predictors consist of random curves \citep[see, e.g.,][]{cardot2005, ferraty2005, cardot2007, chenmuller2012, Kato2012, Tang2014, Yu2016, yao2017, Ma2019, Sang2020, Chaouch2020}. On the other hand, for the QR settings in the context of function-on-scalar regression models, where the response variable involves random curves and predictors are scalar variables \citep[see, e.g.,][]{Kim2007, Wang2009, yang2020, Liu2020}.

As in other cases, it is also important to explore the effects of functional predictors on the functional response (i.e., the function-on-function regression) at the upper and lower tails of the response variable's distribution, not only at the central location. The following two examples can explain the need for QR in function-on-function regression. In a function-on-function regression model, \cite{Antoch2010} predicted averaged weekends' or weekdays' electricity consumption, and provided actual weekdays' electricity consumption is known. However, the energy service providers are generally interested in predicting the 99.99\% of the entire electricity demand distribution to manage the risks adequately, which can easily be done using QR. \cite{LuoQi} analyzed the effects of several air pollutants and meteorological variables on the nitrogen dioxide (NO$_2$) using a conditional mean-based function-on-function interaction regression model. The interactions of the air pollutants and meteorological variables and their behaviors with NO$_2$ may change at different regimes of NO$_2$ concentrations. A conditional mean regression can not explain this association, but QR makes it possible to characterize the effects of pollutants and meteorological variables on the entire conditional distribution of NO$_2$ to understand better the relationships between the functional variables and their interactions. However, to the best of our knowledge, QR has not yet been studied in function-on-function regression models.

This paper extends the standard QR idea into the function-on-function regression model. The proposed model allows for more than one functional predictor in the model, making our proposal attractive when the response variable's conditional distribution needs to be characterized by more than one functional predictor. As in the standard function-on-function regression model, the direct estimation of the proposed model is an ill-posed problem due to the infinite-dimensional nature of the model parameters. The commonly used method in the literature for overcoming this problem is to project the infinite-dimensional model parameters onto a finite-dimensional space using dimension reduction techniques. For this purpose, several techniques have been proposed based on:
\begin{inparaenum}
\item[1)] the general basis expansion functions including $B$-splines, Fourier, wavelet basis, and Gaussian basis \citep[see e.g.,][and references therein]{ramsay1991, ramsay2006, matsui2009, ivanescu2015, chiou2016, BScom},
\item[2)] functional principal component regression (FPCR) \citep[see, e.g.,][]{yao2005, Hall2006, Valderrama, chiou2016, Harjit}, and
\item[3)] functional partial least squares regression (FPLSR) \citep[see, e.g.,][]{PredSap, Reiss2007, PredSc, Delaigle2012, Bande2017, BSenv}.
\end{inparaenum}
The general basis expansion functions may require a large number of basis functions when projecting the regression coefficient function onto a finite-dimensional space, leading to poor estimation and prediction accuracy. The FPCR and FPLSR methods compute orthogonal components from the infinite-dimensional object to produce an approximation in the finite-dimensional space. The FPCR components are computed based on the maximization of the covariance between the functional predictors. In contrast, the FPLSR components are computed by maximizing the covariance between the functional response and functional predictors. Therefore, compared with FPCR, the FPLSR components capture the relevant information with fewer terms, which makes it more preferable than the FPCR \citep[see, e.g.,][for more details]{Reiss2007, Delaigle2012, Bande2017}. In addition, the numerical analyses performed by \cite{Aguilera2010} showed that the FPLSR produces improved regression coefficient estimates compared with FPCR. Therefore, in this study, we consider the FPLSR in the estimation phase of the proposed method.

To estimate the proposed model, we adapt the partial quantile regression (PQR) approach of \cite{Dodge}, which is a QR analog of the standard PLS regression, into the FPLSR. The proposed functional partial quantile regression (FPQR) method uses a partial quantile covariance between the functional random variables to extract the FPQR basis. The extracted basis functions are then used to compute the FPQR components and to estimate the final model. Although the functional random variables belong to an infinite-dimensional space, they are observed in a finite set of discrete-time points in practice. Thus, in the proposed method, the functional forms of the discretely observed random variables are constructed via a finite-dimensional basis function expansion method. The FPQR constructed using the functional random variables is approximated by the multivariate extension of the PQR constructed via the basis expansion coefficients of the response and predictors. The proposed method is an iterative approach, and thus, its finite-sample performance is affected by the number of FPQR basis functions. To this end, a Bayesian information criterion (BIC) is used to determine the optimum number of FPQR components. For a model with multiple predictors, the model's exact form is generally unspecified since the significant variables are unknown in practice. Therefore, a forward stepwise variable selection procedure is used to determine significant functional predictors. Moreover, a nonparametric bootstrap procedure coupled with the proposed method is used to construct pointwise prediction intervals for conditional quantiles of the response variable.

The remaining part of this paper is organized as follows: An overview of PQR is presented in Section~\ref{sec:opqr}. A detailed description of the proposed method is presented in Section~\ref{sec:model}. Several Monte Carlo experiments under different data generation processes and an empirical data example are performed to evaluate the finite-sample performance of the proposed method, and the results are given in Section~\ref{sec:results}. Finally, Section~\ref{sec:conc} concludes the paper, along with some ideas on how the methodology presented here can be further extended.

\section{Overview of PQR} \label{sec:opqr}

Let $Y$ and $\bm{X} = \left[ X_1, \ldots, X_M \right]^\top$ denote a scalar response and $M$-dimensional vector of predictors, respectively. Then, the conditional expectation of $Y$ given $\bm{X}$ (e.g., the linear least squares predictor of $Y$ from $\bm{X}$) is expressed as follows \citep{Dodge}:
\begin{equation}\label{eq:exp}
\text{E}\left[ Y \big\vert \bm{X} \right] = \text{E} \left[ Y \right] + \text{Cov}(Y, \bm{X}) \text{Var}(\bm{X})^{-1} \left( X - \text{E} \left[ X \right] \right),
\end{equation}
which minimizes the squared loss $\text{E}\left[Y - \beta_0 - \bm{\beta}^\top \bm{X} \right]^2$, where $\beta_0$ and $\bm{\beta} = \left[ \beta_1, \ldots, \beta_M \right]^\top$ are the intercept and $M$-dimensional regression coefficient vector, respectively. For a given $\tau \in (0,1)$, let $\text{E}_{\tau} \left[ Y \right]$ and $\text{Cov}_{\tau}(Y, \bm{X})$ denote the quantile expectation of $Y$ and the quantile covariance between $Y$ and $\bm{X}$, respectively, as follows:
\begin{align*}
\text{E}_{\tau} \left[ Y \right] &= \arg~~ \underset{\beta_0}{\inf}~ \text{E} \left[\rho_{\tau}(Y - \beta_0) \right], \\
\text{Cov}_{\tau} (Y, \bm{X})^\top &= \underset{\bm{\beta}}{\arg}~~ \underset{\beta_0, \bm{\beta}}{\inf}~ \text{E}\left[ \rho_{\tau} \left( Y - \beta_0 - \bm{\beta}^\top \text{Var}(\bm{X})^{-1} \left( \bm{X} - \text{E} \left[ \bm{X} \right] \right) \right)  \right],
\end{align*}
where $\rho_{\tau}(u) = u \left\lbrace \tau - \mathbb{1} (u < 0) \right\rbrace$ with the indicator function $\mathbb{1}\{\cdot\}$ denotes the check loss function \citep{Koenker1978, Koenker2005}. Substituting $\text{E}_{\tau} \left[ Y \right]$ and $\text{Cov}_{\tau}(Y, \bm{X})$ for $\text{E} \left[ Y \big\vert \bm{X} \right]$ and $\text{Cov}(Y, \bm{X})$ in~\eqref{eq:exp}, \cite{Dodge} proposed the algorithm of PQR as in Algorithm~\ref{alg:pqr}.

\begin{algorithm}
Centre and scale $\bm{X}$ such that $\text{E} \left[ \bm{X} \right]  = \bm{0}$ and $\text{Var}(\bm{X}) = \bm{1}$.\\
Repeat: \linebreak
\begin{scriptsize}
\textbf{2.1}
\end{scriptsize}
Calculate the direction vector $\bm{c} = \left[c_1, \ldots, c_M \right]^\top$ where $\bm{c}_m = \text{Cov}_{\tau}(Y, X_m)$ (for $m = 1, \ldots, M$), and normalize it so that $\bm{c}^\top \bm{c} = 1$.
\linebreak
\begin{scriptsize}
\textbf{2.2}
\end{scriptsize}
Compute the one-dimensional component vector $T = \bm{c}^\top \bm{X}$ and the least squares predictor $\text{E} \left[ X_m \big\vert T \right] $, and save $T$.
\linebreak
\begin{scriptsize}
\textbf{2.3}
\end{scriptsize}
Update $T$ by replacing $\bm{X}$ by their residuals $X_m - \text{E} \left[ X_m \big\vert T \right] $. \\
Based on the optimum number of component $h$ obtained using an information criterion, retain the components $\left\lbrace T^{(1)}, \ldots, T^{(h)} \right\rbrace $ and obtain the final predictor: $\text{E}_{\tau} \left[ Y \big\vert T^{(1)}, \ldots, T^{(h)} \right] $.
\caption{PQR algorithm \citep{Dodge}}
\label{alg:pqr}
\end{algorithm}

As noted by \cite{Dodge}, the PQR algorithm exactly follows the standard PLS algorithm with two differences:
\begin{inparaenum}
\item[1)] PQR uses quantile covariance and quantile expectation instead of standard covariance and expectation in steps 2.1 and 3 in Algorithm~\ref{alg:pqr}, respectively. However, this difference does not affect the orthogonality of the components, so $ \left\lbrace T^{(1)}, \ldots, T^{(h)} \right\rbrace $ remains mutually orthogonal.
\item[2)] In step 2.3, there is no need to replace $Y$ by its residual when updating the component $T$, as is optional in the standard PLS algorithm and strictly unnecessary due to the orthogonality of $T$ and $X_m - \text{E} \left[ X_m \big\vert T \right] $ \citep[see][]{Dodge}.
\end{inparaenum}
In a nutshell, the PQR includes the features of both the PLS and QR so that it provides efficient prediction by extracting information from the response variable $Y$, as well as it allows to characterize the entire conditional distribution of the response variable.

\section{Function-on-function partial quantile regression}\label{sec:model}

For ease of notation, we introduce the proposed method with a univariate functional regression setting. Let $\left\lbrace \Y_i(t), \X_i(s): i = 1, \ldots, n \right\rbrace$ denote an i.i.d. random sample from a random pair $\left( \Y(t), \X(s) \right)$, where $\Y(t)$ and $\X(s)$, respectively, denote a functional response and a functional predictor, both are the elements of the $\mathcal{L}_2$ separable Hilbert space ($\mathcal{H}$), that is they are square-integrable and real-valued functions defined on the closed and bounded intervals $t \in \mathcal{I}$ and $s \in \mathcal{S}$. Without loss of generality, we assume that the functional response and functional predictor are zero-mean processes so that $\text{E} \left[ \Y(t) \right]  = \text{E} \left[ \X(s) \right] = 0$ and $\forall t,s \in \left[ 0,1 \right] $.

The function-on-function regression model is expressed as follows:
\begin{equation}\label{eq:fof}
\text{E} \left[ \Y_i(t) \vert \X_i(t) \right] = \int_0^1 \X_i(s) \beta(s,t) ds,
\end{equation}
where $\beta(s,t)$ is a smooth bivariate regression coefficient function that measures the effect of $\X_i(s)$ on the conditional mean of $\Y_i(t)$. We define a more general alternative to the function-on-function regression model to exhibit a more comprehensive description of the entire distribution of the functional response $\Y(t)$. For a given $\tau \in (0,1)$, we denote the $\tau$\textsuperscript{th} conditional quantile of the functional response given the functional predictor by $Q_{\tau}\left[ \Y(t) \big\vert \X(s) \right]$, as follows:
\begin{equation}\label{eq:cq}
Q_{\tau}\left[ \Y_i(t) \big\vert \X_i(s) \right] = \int_0^1 \X_{i}(s) \beta_{\tau}(s,t) ds,
\end{equation}
where the regression coefficient function $\beta_{\tau}(s,t)$ measures the effect of $\X_i(s)$ on the $\tau$\textsuperscript{th} quantile of $\Y_i(t)$. In the standard function-on-function regression model~\eqref{eq:fof}, the regression coefficient function is assumed to be fixed. On the other hand, the regression coefficient function $\beta_{\tau}(s,t)$ in~\eqref{eq:cq} varies with respect to $\tau$ over the function support, and thus, it allows to characterize the entire conditional distribution of the functional response.

To estimate $\beta_{\tau}(s,t)$ and obtain the PQR predictor of $Q_{\tau}\left[ \Y_i(t) \big\vert \X_i(s) \right]$ in~\eqref{eq:cq}, we propose a FPQR method, which is an adaptation of the PQR idea of \cite{Dodge} into the FPLS regression approach discussed by \cite{PredSc} and \cite{BSenv}. The FPLS components of Model~\eqref{eq:fof} are obtained by maximizing the squared covariance between the functional response and functional predictor, where the least-squares loss function is used to optimize the covariance operator, as follows:
\begin{align}
&\arg~~\underset{\begin{subarray}{c}
  p \in \mathcal{L}_2[0,1],~ \Vert p \Vert_{\mathcal{L}_2[0,1]} = 1 \\
  w \in \mathcal{L}_2[0,1],~ \Vert w \Vert_{\mathcal{L}_2[0,1]} = 1
  \end{subarray}}{\max} \text{Cov}^2 \left( \int_0^1 \Y(t) p(t) dt, ~ \int_0^1 \X(s) w(s) ds \right), \label{eq:cov_cv} \\
\Longleftrightarrow &  \arg~~\underset{\begin{subarray}{c}
  p \in \mathcal{L}_2[0,1],~ \Vert p \Vert_{\mathcal{L}_2[0,1]} = 1 \\
  w \in \mathcal{L}_2[0,1],~ \Vert w \Vert_{\mathcal{L}_2[0,1]} = 1
  \end{subarray}}{\min} \text{E}^2 \left[ \varphi \left( \int_0^1 \Y(t) p(t) dt - \int_0^1 \X(s) w(s) ds \right) \right], \label{eq:cov_ls}
\end{align}
where $p(t)$ and $w(s)$ are the weight functions and $\varphi(u) = u^2$ is the least-squares loss function. The usual covariance in~\eqref{eq:cov_cv} is used to predict the mean value of the response variable for given predictors. On the other hand, to predict the quantiles of the response, the covariance operator needs to be redefined. For this purpose, i.e., to obtain the FPQR components of the proposed model~\eqref{eq:cq}, we define a quantile covariance, denoted by $\text{Cov}_{\tau} \left( \cdot, \cdot \right) $, by replacing the least-squares loss function in~\eqref{eq:cov_ls} by the quantile loss function, i.e., $\rho_{\tau}(u) = u \left\lbrace \tau - \mathbb{1} (u < 0) \right\rbrace$. Since the optimizations with the quantile loss function at different $\tau$ levels lead to different results, the quantile covariance depends on the $\tau$ level.

To start with, let $\mathcal{C}_{y x}^{\tau}$ denote the quantile covariance between the functional random variables $y(\cdot)$ and $x(\cdot)$, which evaluates the contribution of $x(\cdot)$ to the $\tau\textsuperscript{th}$ quantile of $y(\cdot)$. Let $\mathcal{C}_{\Y \X}^{(\tau)}$ and $\mathcal{C}_{\X \Y}^{(\tau)}$ denote the cross-quantile-covariance operators as follows:
\begin{align*}
\mathcal{C}_{\Y \X}^{\tau} &= \mathcal{L}_2[0,1] \rightarrow \mathcal{L}_2[0,1], \qquad f \rightarrow g = \int_0^1 \text{Cov}_{\tau} \left[ \Y(t), \X(s) \right] f(t) dt,\\
\mathcal{C}_{\X \Y}^{\tau} &= \mathcal{L}_2[0,1] \rightarrow \mathcal{L}_2[0,1], \qquad g \rightarrow f = \int_0^1 \text{Cov}_{\tau} \left[ \Y(t), \X(s) \right]  g(s) ds.
\end{align*} 
Then, from the $\mathcal{L}_2$ continuity of $\Y(t)$ and $\X(s)$, $\mathcal{U} = \mathcal{C}_{\X \Y}^{\tau} \circ \mathcal{C}_{\Y \X}^{\tau}$ and $\mathcal{V} = \mathcal{C}_{\Y \X}^{\tau} \circ \mathcal{C}_{\X \Y}^{\tau}$ are defined as self-adjoint, positive, and compact operators, whose spectral analyses lead to a countable set of positive eigenvalues $\lambda_{\tau}$ associated to orthonormal eigenfunctions $w_{\tau} \in \mathcal{L}_[0,1]$ as a solution of
\begin{equation}\label{eq:sol}
\mathcal{U} w_{\tau} = \lambda_{\tau} w_{\tau},
\end{equation}
where $\int_0^1 w_{\tau}(s) w_{\tau}^\top(s) ds = 1$ \citep[see, e.g.,][for more information]{PredSap}. We consider the following optimization problem to obtain the FPQR components of Model~\eqref{eq:cq}:
\begin{align}
& \arg~~\underset{\begin{subarray}{c}
  p_{\tau} \in \mathcal{L}_2[0,1],~ \Vert p_{\tau} \Vert_{\mathcal{L}_2[0,1]} = 1 \\
  w_{\tau} \in \mathcal{L}_2[0,1],~ \Vert w_{\tau} \Vert_{\mathcal{L}_2[0,1]} = 1
  \end{subarray}}{\max} \text{Cov}_{\tau}^2 \left( \int_0^1 \Y(t) p_{\tau}(t) dt, ~ \int_0^1 \X(s) w_{\tau}(s) ds \right), \label{eq:optc} \\
  \Longleftrightarrow &  \arg~~\underset{\begin{subarray}{c}
  p_{\tau} \in \mathcal{L}_2[0,1],~ \Vert p_{\tau} \Vert_{\mathcal{L}_2[0,1]} = 1 \\
  w_{\tau} \in \mathcal{L}_2[0,1],~ \Vert w_{\tau} \Vert_{\mathcal{L}_2[0,1]} = 1
  \end{subarray}}{\min} \text{E}^2 \left[\rho_{\tau} \left( \int_0^1 \Y(t) p_{\tau}(t) dt - \int_0^1 \X(s) w_{\tau}(s) ds \right) \right], \label{eq:opte}
\end{align}
where $p_{\tau}(t)$ and $w_{\tau}(s)$ are the eigenfunctions for quantile level $\tau$ associated with the largest eigenvalue of $\mathcal{V}$ and $\mathcal{U}$, respectively.

Let $w_{\tau}^{(1)}(s)$ denote the eigenfunction of $\mathcal{U}$ for quantile level $\tau$ associated with the largest eigenvalue, denoted by $\lambda_{\tau, \max}$, as follows:
\begin{equation*}
\mathcal{U} w_{\tau}^{(1)}(s) = \lambda_{\tau,\max} w_{\tau}^{(1)}(s).
\end{equation*}
Then, the first FPQR component, denoted by $T_{\tau}^{(1)}$, is obtained using the functional linear regression as follows:
\begin{equation*}
T_{\tau}^{(1)} = \int_0^1 w_{\tau}^{(1)}(s) \X(s) ds.
\end{equation*}
Similar to FPLS, the proposed FPQR method uses an iterative procedure to determine subsequent FPQR components. Let $h = 1, \ldots, H$ be the iteration step. Let $\X^{(0)}(s) = \X(s)$, and 
\begin{equation*}
\X^{(h)}(s) = \X^{(h-1)}(s) - \text{E}\left[ \X^{(h-1)}(s) \big\vert T_{\tau}^{(h)}\right], 
\end{equation*}
where $\text{E}\left[\X^{(h-1)}(s) \big\vert T_{\tau}^{(h)}\right]$ = $\text{E}\left[ \X^{(h-1)}(s)\right] -\text{Cov}\left[ \X^{(h-1)}(s), T_{\tau}^{(h)}\right]$ $\text{Var} \left( T_{\tau}^{(h)} \right)^{-1} \left( T_{\tau}^{(h)} - \text{E} \left[ T_{\tau}^{(h)} \right] \right)$ denotes the residuals of the functional linear regression of $\X^{(h-1)}(s)$ on $T_{\tau}^{(h)}$. Then, at step $h$, $w_{\tau}^{(h)}(s)$ is obtained as the solution of 
\begin{align*}
w_{\tau}^{(h)} &= \arg~~\underset{\begin{subarray}{c}
  p_{\tau} \in \mathcal{L}_2[0,1],~ \Vert p_{\tau} \Vert_{\mathcal{L}_2[0,1]} = 1 \\
  w_{\tau} \in \mathcal{L}_2[0,1],~ \Vert w_{\tau} \Vert_{\mathcal{L}_2[0,1]} = 1
  \end{subarray}}{\max} \text{Cov}_{\tau}^2 \left( \int_0^1 \Y(t) p_{\tau}(t) dt, ~ \int_0^1 \X^{(h-1)}(s) w_{\tau}(s) ds \right), \\
 & \Longleftrightarrow  \arg~~\underset{\begin{subarray}{c}
  p_{\tau} \in \mathcal{L}_2[0,1],~ \Vert p_{\tau} \Vert_{\mathcal{L}_2[0,1]} = 1 \\
  w_{\tau} \in \mathcal{L}_2[0,1],~ \Vert w_{\tau} \Vert_{\mathcal{L}_2[0,1]} = 1
  \end{subarray}}{\min} \text{E}^2 \left[\rho_{\tau} \left( \int_0^1 \Y(t) p_{\tau}(t) dt - \int_0^1 \X^{(h-1)}(s) w_{\tau}(s) ds \right) \right],
\end{align*}
which is the eigenfunction of $\mathcal{U}^{(h-1)} = \mathcal{C}_{\X^{(h-1)} \Y}^{\tau} \circ \mathcal{C}_{\Y \X^{(h-1)}}^{\tau}$, where $\mathcal{C}_{\X^{(h-1)} \Y}^{\tau}$ is the cross-quantile-covariance operator of $\bm{\X}^{(h-1)}(s)$ and $\Y(t)$, that is
\begin{equation*}
\mathcal{U}^{(h-1)} w_{\tau}^{(h)}(s) = \lambda_{\tau,\max} w_{\tau}^{(h)}(s).
\end{equation*}
Then, $h\textsuperscript{th}$ FPQR component, $T_{\tau}^{(h)}$, is obtained using the functional linear regression as follows:
\begin{equation*}
T_{\tau}^{(h)} = \int_0^1 w_{\tau}^{(h)}(s) \X^{(h-1)}(s) ds.
\end{equation*}

Let $\left\lbrace T_{\tau}^{(1)}, \ldots, T_{\tau}^{(H)} \right\rbrace$ denote the retained FPQR components after $H$ iterations. In the last step of the FPLS method, the PLS approximations of the function-on-function mean regression model are obtained by conducting an ordinary linear regression of the response on the retained components. On the other hand, the proposed method is finalized by conducting a QR model of the response on the retained FPQR components, $\text{E} \left[\rho_{\tau} \left(  \Y(t) \big\vert T_{\tau}^{(1)}, \ldots, T_{\tau}^{(h)} \right) \right] $, to obtain the PQR approximations of the model parameter and the quantile of the response.

In summary, compared with the FPLS, the proposed FPQR follows a similar procedure with two differences. First, the FPQR uses quantile covariance when computing the components instead of the usual covariance. Second, in the last step of the proposed method, the final PQR approximations of Model~\eqref{eq:cq} are obtained via a QR of the response on the retained FPQR components.

\subsection{PQR for basis expansion of the functional QR}\label{sec:bfe}

Although the functional random variables $\left[ \Y(t), \X(s) \right]$ naturally belong to an infinite-dimensional space, the randomly observed sample curves are observed in the finite sets of time points such that $\left\lbrace \Y_i(t_{l}), \X_i(s_r): l = 1, \ldots, L,~ r = 1, \ldots,R \right\rbrace$. Thus, in a functional regression method, the functional forms of the random variables are first approximated from the discretely observed data points before fitting the model. To this end, several approaches, such as nonparametric smoothing of functions \citep[see, e.g.,][]{ferraty2006} and basis function expansion \citep[see, e.g.,][]{ramsay2006} have been proposed. In this paper, we consider basis function expansion method to construct the functional forms of the random variables. In summary, for a sufficiently large number of basis functions $K$, a function $y(t)$ is projected into the finite-dimensional space as a linear combinations of basis functions, $\phi_k(s)$, and their corresponding coefficients, $b_k$, as follows:
\begin{equation*}
y(t) \approx \sum_{k=1}^K b_k \phi_k(t).
\end{equation*}
For this purpose, several basis function expansion methods, such as, radial, wavelet, $B$-spline, and Fourier, have been proposed, \citep[see, e.g,][]{ramsay2006}. In this paper, we consider $B$-spline basis expansion method since it is one of the most commonly used basis expansion method in practice. Let $\left\lbrace \phi_k(t): k = 1, \ldots, K_{\Y} \right\rbrace$ and $\left\lbrace \psi_j(s): j = 1, \ldots, K_{\X} \right\rbrace$ denote $K_{\Y}$ and $K_{\X}$ dimensional basis functions to project $\Y(t)$ and $\X(s)$, respectively. Then, the functional response and functional predictor are expressed in the basis expansion form as follows:
\begin{align*}
\Y(t) &\approx \sum_{k=1}^{K_{\Y}} b_k \phi_k(t) = \bm{b}^\top \bm{\Phi}(t), \\
\X(s) &\approx \sum_{j=1}^{K_{\X}} a_j \psi_j(s) = \bm{a}^\top \bm{\Psi}(s),
\end{align*}
where $\bm{b} = \left[ b_1, \ldots, b_{K_{\Y}} \right]^\top$ and $\bm{a} = \left[ a_1, \ldots, a_{K_{\X}} \right]^\top$ are the vectors of basis expansion coefficients.

From~\eqref{eq:sol}, the FPQR eigenfunctions can be expressed in terms of basis expansion function as follows:
\begin{align*}
p_{\tau}(t) &\approx \sum_{k=1}^{K_{\Y}} p_k \phi_k(t) = \bm{p}^\top \bm{\Phi}(t), \\
w_{\tau}(s) &\approx \sum_{j=1}^{K_{\X}} w_j \psi_j(s) = \bm{w}^\top \bm{\Psi}(s),
\end{align*}
where $\bm{p} = \left[ p_1, \ldots, p_{K_{\Y}} \right]^\top$ and $\bm{w} = \left[ w_1, \ldots, w_{K_{\X}} \right]^\top$ are the vectors of basis expansion coefficients of $p_{\tau}(t)$ and $w_{\tau}(s)$, respectively. Thus, the regression coefficient function $\beta_{\tau}(s,t)$ can also be exppressed in the bases $\bm{\Phi}(t)$ and $\bm{\Psi}(s)$ as follows:
\begin{equation}
\beta_{\tau}(s,t) = \sum_{j=1}^{K_{\X}} \sum_{k=1}^{K_{\Y}} \psi_j(s) \beta_{jk} \phi_k(t) = \bm{\Psi}^\top(s) \bm{\beta} \bm{\Phi}(t),
\end{equation}
where $\bm{\beta} = \left( \beta_{jk} \right)_{jk}$ is a $K_{\X} \times K_{\Y}$ matrix of basis expansion coefficients. Moreover, the cross-quantile-covariance operators can be expressed in terms of basis expansions of $\Y(t)$ and $\X(s)$ as follows:
\begin{align*}
\mathcal{C}_{\Y \X}^{(\tau)} &= \mathcal{L}_2[0,1] \rightarrow \mathcal{L}_2[0,1], \qquad f \rightarrow g = \left( \Sigma_{\bm{a} \bm{b}}^{\tau}\right)^\top \bm{\Psi} f,\\
\mathcal{C}_{\X \Y}^{(\tau)} &= \mathcal{L}_2[0,1] \rightarrow \mathcal{L}_2[0,1], \qquad g \rightarrow f = \Sigma_{\bm{a} \bm{b}}^{\tau} \bm{\Phi} g,
\end{align*}
where $\Sigma_{\bm{a} \bm{b}}^{\tau}$ is the cross-quantile-covariance matrix between the basis expansion coefficients $\bm{a}$ and $\bm{b}$ and $\bm{\Psi} = \int_0^1 \bm{\Psi}(s) \bm{\Psi}^\top(s) ds$ and $\bm{\Phi} = \int_0^1 \bm{\Phi}(t) \bm{\Phi}^\top(t) dt$ are the $K_{\X} \times K_{\X}$ and $K_{\Y} \times K_{\Y}$ inner product matrices, respectively. 

Based on the basis function expansions of the cross-quantile-covariance operators,~\eqref{eq:sol} can be rewritten in terms of bais expansion coefficient so that the first PQR eigenfunction can be expressed as follows:
\begin{equation}\label{eq:solb}
\Sigma_{\bm{a} \bm{b}}^{\tau} \bm{\Phi} \left( \Sigma_{\bm{a} \bm{b}}^{\tau}\right)^\top \bm{\Psi} w_{\tau}^{(1)} = \lambda_{\tau, \max} w_{\tau}^{(1)}.
\end{equation}
Let us now consider the decompositions $\bm{\Phi} = \left( \bm{\Phi}^{1/2} \right) \left( \bm{\Phi}^{1/2} \right)^\top$ and $\bm{\Psi} = \left( \bm{\Psi}^{1/2} \right) \left( \bm{\Psi}^{1/2} \right)^\top$. In addition, let us consider the following equality:
\begin{equation*}
\left( w_{\tau}^{(1)} \right)^\top \bm{\Psi} w_{\tau}^{(1)} = \left( w_{\tau}^{(1)} \right)^\top \left( \bm{\Phi}^{1/2} \right) \left( \bm{\Phi}^{1/2} \right)^\top w_{\tau}^{(1)} = \left( \widetilde{w}_{\tau}^{(1)} \right)^\top \widetilde{w}_{\tau}^{(1)},
\end{equation*}
where $\widetilde{w}_{\tau}^{(1)} = \left( \bm{\Phi}^{1/2} \right)^\top w_{\tau}^{(1)}$. Accordingly,~\eqref{eq:solb} can be expressed as follows:
\begin{equation}\label{eq:solf1}
\left( \bm{\Psi}^{1/2} \right)^\top \Sigma_{\bm{a} \bm{b}}^{\tau} \bm{\Phi}^{1/2} \left( \bm{\Phi}^{1/2} \right)^\top \left( \Sigma_{\bm{a} \bm{b}}^{\tau}\right)^\top \bm{\Psi}^{1/2} \widetilde{w}_{\tau}^{(1)} = \lambda_{\tau, \max} \widetilde{w}_{\tau}^{(1)}.
\end{equation}

At step $h$, the $h\textsuperscript{th}$ PQR component can be obtained in terms of the associated eigenfunction by solving
\begin{equation}\label{eq:solf2}
\left( \bm{\Psi}^{1/2} \right)^\top \Sigma_{\bm{a}^{(h-1}) \bm{b}}^{\tau} \bm{\Phi}^{1/2} \left( \bm{\Phi}^{1/2} \right)^\top \left( \Sigma_{\bm{a}^{(h-1}) \bm{b}}^{\tau}\right)^\top \bm{\Psi}^{1/2} \widetilde{w}_{\tau}^{(h)} = \lambda_{\tau, \max} \widetilde{w}_{\tau}^{(h)},
\end{equation}
where  $\Sigma_{\bm{a}^{(h-1}) \bm{b}}^{\tau}$ denotes the cross-quantile-covariance matrix between the basis expansion coefficient vectors of $\X^{(h-1)}(s)$ and $\Y(t)$, respectively denoted by $\bm{a}^{(h-1)}$ and $\bm{b}$ \citep[see e.g.,][for more details]{Aguilera2019}. From~\eqref{eq:solf1} and~\eqref{eq:solf2} and the fact that $\left( \bm{\Psi}^{1/2} \right)^\top \Sigma_{\bm{a} \bm{b}}^{\tau} \bm{\Phi}^{1/2}$  is the cross-quantile-covariance matrix between $\bm{b}^\top \bm{\Phi}^{1/2}$ and $\bm{a}^\top \bm{\Psi}^{1/2}$, it can be concluded that PFQR model is equivalent to multivariate PQR of $\bm{b}^\top \bm{\Phi}^{1/2}$ on $\bm{a}^\top \bm{\Psi}^{1/2}$.

Let $\bm{\Omega} = \bm{b}^\top \bm{\Phi}^{1/2}$ and $\bm{Z} = \bm{a}^\top \bm{\Psi}^{1/2}$. Then, in the finite-dimensional space of basis expansion coefficients, the $\tau$\textsuperscript{th} conditional quantile of the functional response $\Y(t)$ given the functional predictor $\X(s)$ in~\eqref{eq:cq} is expressed as follows:
\begin{equation*}
Q_{\tau}\left[ \bm{\Omega} \big\vert \bm{Z} \right] = \bm{Z} \bm{\Theta}_{\tau},
\end{equation*}
where $\bm{\Theta}_{\tau}$ denotes the coefficient matrix, and it can be estimated by minimizing the check loss function as follows:
\begin{equation*}
\bm{\widehat{\Theta}}_{\tau} = \underset{\bm{\Theta}_{\tau}}{\argmin} \left[ \sum_{i=1}^n \rho_{\tau} \left( \bm{\Omega}_i - \bm{Z}_i \bm{\Theta}_{\tau} \right) \right].
\end{equation*}
To obtain an estimate for $\bm{\Theta}_{\tau}$, we propose a multivariate extension of the PQR method of \cite{Dodge} in Algorithm~\ref{alg:mpqr}.

\begin{algorithm}[ht]
Centre and scale $\bm{Z}$ such that $\text{E}(\bm{Z}) = \bm{0}$ and $\text{Var}(\bm{Z}) = \bm{1}$.\\
Repeat: \linebreak
\begin{scriptsize}
\textbf{2.1}
\end{scriptsize}
Calculate the $K_{\X} \times K_{\Y}$ dimensional direction matrix $\bm{c} = \text{Cov}_{\tau}(\bm{\Omega}, \bm{Z})$ with $\bm{c} = \left[ \bm{c}_1, \ldots, \bm{c}_{K_{\Y}} \right]^\top$ and $\bm{c}_k = \left[ c_{k1}, \ldots, c_{k K_{\X}} \right]^\top$, and normalize it so that $\bm{c}_k^\top \bm{c}_k = 1$ for $k = 1, \ldots, K_{\Y}$.
\linebreak
\begin{scriptsize}
\textbf{2.2}
\end{scriptsize}
Compute the $(n \times K_{\Y})$-dimensional component matrix $\bm{T} = \bm{c}^\top \bm{Z}$ and the least squares predictor $\text{E} \left[ \bm{Z} \big\vert \bm{T} \right] $ such that:
\begin{equation*}
\text{E} \left[ \bm{Z} \big\vert \bm{T} \right] = \text{E} \left[ \bm{Z} \right]  + \text{Cov}(\bm{Z}, \bm{T}) \text{Var}(\bm{T})^{-1} \left( \bm{T} - \text{E} \left[ \bm{T} \right]  \right),
\end{equation*}
and save $T$.
\linebreak
\begin{scriptsize}
\textbf{2.3}
\end{scriptsize}
Update $\bm{T}$ by replacing $\bm{Z}$ by their residuals $\bm{Z} - \text{E} \left[ \bm{Z} \big\vert \bm{T} \right] $. \\
Based on the optimum number of component $h$ obtained using an information criterion, retain the components $\left\lbrace \bm{T}^{(1)}, \ldots, \bm{T}^{(h)} \right\rbrace $ and obtain the final predictor: $\text{E}_{\tau} \left[ \bm{\Omega} \big\vert \bm{T}^{(1)}, \ldots, \bm{T}^{(h)} \right]$.
\caption{Multivariate PQR algorithm}
\label{alg:mpqr}
\end{algorithm}

Let $\widehat{\bm{\Theta}}_{\tau}^{(h)}$ denote the estimated regression coefficient of $\bm{\Theta}_{\tau}^{(h)}$, obtained after $h$ iterations using the multivariate PQR algorithm~\ref{alg:mpqr}. Then, the the FPQR approximation of $\tau$\textsuperscript{th} conditional quantile of the functional response $\Y(t)$ given the functional predictor $\X(s)$ is given by
\begin{align*}
\widehat{Q}_{\tau}\left[ \bm{\Omega} \big\vert \bm{Z} \right] &= \bm{a}^\top \bm{\Psi}^{1/2} \widehat{\bm{\Theta}}_{\tau}^{(h)} = \bm{a}^\top \bm{\Psi}^{1/2} \left[ \bm{\Psi}^{1/2} \left( \bm{\Psi}^{1/2} \right)^{-1} \right] \widehat{\bm{\Theta}}_{\tau}^{(h)} \\
\widehat{Q}_{\tau}\left[ \Y_i(t) \big\vert \X_i(s) \right] &= \int_0^1 \X(t) \widehat{\beta}_{\tau}^{(h)}(s,t) ds,
\end{align*}
where
\begin{equation*}
\widehat{\beta}_{\tau}^{(h)}(s,t) = \left[ \left( \bm{\Psi}^{1/2} \right)^{-1} \widehat{\bm{\Theta}}_{\tau}^{(h)} \left( \bm{\Phi}^{1/2} \right)^{-1} \right] \bm{\Phi}(t) \bm{\Psi}(s).
\end{equation*}

Based on the results presented above, the infinite-dimensional problems of estimating the regression coefficient function $\beta_{\tau}(s,t)$ and $\tau$\textsuperscript{th} conditional quantile of the functional response $\Y(t)$ are reduced to a multivariate finite-dimensional PQR problem. The coefficient matrix of the multivariate PQR between the metrics $\bm{\Phi}$ and $\bm{\Psi}$ in the spaces of expansion coefficients $\bm{b}$ and $\bm{a}$ can be estimated using the available \texttt{R} package ``\texttt{quantreg}'' \citep{quantreg}.

\subsection{Determination of the optimum number of FPQR components}

The proposed FPQR is an iterative procedure. Its finite-sample performance depends on the number of components $h$ used to obtain the regression coefficient estimate and the final predictor of the response variable's conditional quantile. To determine the optimum number of FPQR components, we consider the BIC. Let us denote by $\mathcal{J} = \left\lbrace h \vert h = 1, 2, \ldots \right\rbrace$ all possible models. Then, we assume that there exists a true model for the FPQR model associated with $h_0 \in \mathcal{J}$:
\begin{equation*}
Q_{\tau}\left[ \Y_i(t) \big\vert \X_i(s) \right] = \int_0^1 \X_{i}(s) \beta_{\tau}^{(h_0)}(s,t) ds.
\end{equation*}
For each $h$, the estimated regression coefficient function is given by
\begin{equation*}
\widehat{\beta}_{\tau}^{(h)}(s,t) = \underset{\begin{subarray}{c}
  \beta_{\tau}^{(h)}(s,t)
  \end{subarray}}{\argmin} \left[ \sum_{i=1}^n \rho_{\tau} \left( \Y_i(t) - \int_0^1 \X_i(s) \beta_{\tau}^{(h)}(s,t) ds \right) \right]. 
\end{equation*}
Following the definition of BIC in \cite{Schwarz}, we obtain the following BIC for the FPQR:
\begin{equation*}
\text{BIC}(h) = \ln \bigg \Vert \left[ \sum_{i=1}^n \rho_{\tau} \left( \Y_i(t) - \int_0^1 \X_i(s) \widehat{\beta}_{\tau}^{(h)}(s,t) ds \right) \right] \bigg \Vert_{\mathcal{L}_2} + h \ln(n).
\end{equation*}

To determine the optimum $h$, we buid $h = 1, \ldots, H$ different FPQR models. Then, the optimum number $h$, denoted by $\widetilde{h}$, corresponds to $\widetilde{h} = \underset{\begin{subarray}{c} h \end{subarray}}{\argmin} ~\text{BIC}(h)$.

\subsection{Multiple FPQR model}

In this section, we extend the function-on-function linear QR model to that with multiple functional predictor case and summarize how the proposed FPQR method is used to estimate this model. Let $\left\lbrace \X_1(s), \ldots, \X_M(s) \right\rbrace$ denote $M$ set of functional predictors with $\X_m(s) \in \mathcal{L}_2[0,1],~\forall~m = 1, \ldots, M$. Denote by $\bm{\X}(t) = \left[ \X_1(s), \ldots, \X_M(s) \right]^\top \in \mathcal{H}^M = \mathcal{L}_2^M[0,1]$ the vector of $M$-dimensional vector-valued functions in a Hilbert space. We postulate that $\bm{\X}(s)$ is a $\mathcal{L}_2$ continuous stochastic process, which implicates the $\mathcal{L}_2$ continuity of each component of $\bm{\X}(s)$. Then, the function-on-function linear QR model in~\eqref{eq:cq} is extended to
\begin{align}
Q_{\tau}\left[ \Y_i(t) \big\vert \bm{\X}_i(s) \right] &= \sum_{m=1}^M \int_0^1 \X_{im}(s) \beta_{\tau m}(s,t) ds, \nonumber \\
&= \int_0^1 \bm{\X}_{i}^\top(s) \bm{\beta}_{\tau}(s,t) ds, \label{eq:mcq}
\end{align}
where $\bm{\X}_{i}(s) = \left[ \X_{i1}(s), \ldots, \X_{iM}(s) \right]^\top$ and $\bm{\beta}_{\tau}(s,t) = \left[ \beta_{\tau 1}(s,t), \ldots, \beta_{\tau M}(s,t) \right]^\top \in \mathcal{L}_2^M[0,1]$ denotes the vector of regression coefficient functions.

The proposed FPQR method can be used to estimate the multiple function-on-function linear QR model in a similar way as presented in Section~\ref{sec:model} by extending the cross-quantile covariance operators to $M$-dimensional case. Let $\mathcal{C}_{\Y \bm{\X}}^{\tau}$ and $\mathcal{C}_{\bm{\X} \Y}^{\tau}$ respectively denote the cross-quantile-covariance operator evaluating the contribution of $M$-variate functional predictor $\bm{\X}(s)$ to the functional response $\Y(t)$ and its adjoint as follows:
\begin{align*}
\mathcal{C}_{\Y \bm{\X}}^{\tau} &= \mathcal{L}_2^M[0,1] \rightarrow \mathcal{L}_2[0,1], \qquad f \rightarrow g = \int_0^1 \text{Cov}_{\tau} \left( \Y(t), \bm{\X}(s) \right) \bm{f}(t) dt,\\
\mathcal{C}_{\bm{\X} \Y}^{\tau} &= \mathcal{L}_2[0,1] \rightarrow \mathcal{L}_2^M[0,1], \qquad g \rightarrow f = \int_0^1 \text{Cov}_{\tau} \left( \Y(t), \bm{\X}(s) \right) g(s) ds.
\end{align*}
Then, the FPQR components of Model~\eqref{eq:mcq} can be obtained by iteratively maximizing the squared covariance between $\Y(t)$ and $\bm{\X}(s)$ as follows:
\begin{align*}
& \arg~~\underset{\begin{subarray}{c}
  p_{\tau} \in \mathcal{L}_2[0,1],~ \Vert p_{\tau} \Vert_{\mathcal{L}_2[0,1]} = 1 \\
  \bm{w}_{\tau} \in \mathcal{L}_2^M[0,1],~ \Vert \bm{w}_{\tau m} \Vert_{\mathcal{L}_2[0,1]} = 1,~ \forall~m=1,\ldots,M
  \end{subarray}}{\max} \text{Cov}_{\tau}^2 \left( \int_0^1 \Y(t) p_{\tau}(t) dt, ~ \int_0^1 \bm{\X}^\top(s) \bm{w}_{\tau}(s) ds \right), \\
  \Longleftrightarrow &  \arg~~\underset{\begin{subarray}{c}
  p_{\tau} \in \mathcal{L}_2[0,1],~ \Vert p_{\tau} \Vert_{\mathcal{L}_2[0,1]} = 1 \\
  \bm{w}_{\tau} \in \mathcal{L}_2^M[0,1],~ \Vert \bm{w}_{\tau m} \Vert_{\mathcal{L}_2[0,1]} = 1,~ \forall~m=1,\ldots,M
  \end{subarray}}{\min} \text{E}^2 \left[\rho_{\tau} \left( \int_0^1 \Y(t) p_{\tau}(t) dt - \int_0^1 \bm{\X}^\top(s) \bm{w}_{\tau}(s) ds \right) \right],
\end{align*}
where $p_{\tau}(t)$ and $\bm{w}_{\tau}(s) = \left[ w_{\tau 1}(s), \ldots, w_{\tau M}(s) \right]^\top \in \mathcal{L}_2^M[0,1]$ are the eigenfunctions for quantile level $\tau$ associated with the largest eigenvalue of $\mathcal{V} = \mathcal{C}_{\Y \bm{\X}}^{\tau} \circ \mathcal{C}_{\bm{\X} \Y}^{\tau}$ and $\mathcal{U} = \mathcal{C}_{\bm{\X} \Y}^{\tau} \circ \mathcal{C}_{\Y \bm{\X}}^{\tau}$, respectively.

Similar to Section~\ref{sec:bfe}, using the basis function expansion of the functional variables, it can easily be shown that the FPQR of $\Y(t)$ on $\bm{\X}(s)$ is equivalent to multivariate PQR of $\bm{b}^\top \bm{\Phi}^{1/2}$ on $\bm{A}^\top \bm{\Psi}^{1/2}$. Herein, $\bm{A}$ and $\bm{\Psi}$ denote the basis expansion coefficients of $M$-variate functional predictor $\bm{\X}(s)$. If we assume that each curve $\X_{im}$ for $i = 1, \ldots, n$ and $m = 1, \ldots, M$ is approximated by a basis expansion
\begin{equation*}
\X_{im}(s) \approx \sum_{j=1}^{K_m} a_{imj} \psi_{mj}(s).
\end{equation*}
Then, we have the following matrix representation:
\begin{equation*}
\bm{\X}_i(s) \approx \bm{\Psi}(s) \bm{a}_i,
\end{equation*}
where $\bm{a}_i = \left[ a_{i11}, \ldots, a_{i1K_1}, a_{i21}, \ldots, a_{i2K_2}, \ldots, a_{iM1}, \ldots, a_{iMK_M} \right]^\top$ and
\begin{equation*}
\bm{\Psi}(s) = \begin{bmatrix} 
\bm{\Psi}_1^\top(s) & \bm{0} & \cdots & \bm{0} \\
\bm{0} & \bm{\Psi}_2^\top(s) & \cdots & \bm{0} \\
\cdots & \cdots & \cdots & \cdots \\
\bm{0} & \bm{0} & \cdots & \bm{\Psi}_M^\top(s)
\end{bmatrix},
\end{equation*}
where $\bm{\Psi}_m(s) = \left[ \psi_{m1}(s), \ldots, \psi_{m K_m}(s) \right]^\top$ for $m = 1, \ldots, M$. Let $\bm{A}$ denote the $n \times \sum_{m=1}^M K_m$-dimensional matrix with row entries $\bm{a}_i^\top$. Then, we have the following basis expansion approximation for $\bm{\X}(t)$:
\begin{equation*}
\bm{\X}(s) \approx \bm{A} \bm{\Psi}^\top(s).
\end{equation*}
Finally, we note that $\bm{\Psi} = \int_0^1 \bm{\Psi(s)}$ is a symmetric block-diagonal $\sum_{m=1}^M K_m \times \sum_{m=1}^M K_m$-matrix of the inner products between the basis functions \citep[see e.g.,][]{Julien2014}.

\subsection{Variable selection procedure}\label{sec:vs}

When considering the multiple function-on-function linear QR model~\eqref{eq:mcq}, the vector of functional predictors $\bm{\X}(s)$ may include a great number of functional predictors. Still, not all of them may significantly affect the conditional distribution of the response variable $\Y(t)$. In such cases, a variable selection procedure is needed to determine the variables that significantly affect the response variable's conditional distribution. To this end, we consider a forward stepwise variable selection procedure along with the extension of BIC presented by \cite{Lee2014} to function-on-function linear QR model. Let $\mathcal{M} = \left\lbrace m_1, \ldots, m_d \right\rbrace \subset \left\lbrace 1, \ldots, M \right\rbrace$ denote a candidate model including functional predictors $\left\lbrace \X_{m_1}(s), \ldots, \X_{m_d}(s) \right\rbrace$ and let $\bm{\X}^{\mathcal{M}}(s) = \left[ \X_{m_1}(s), \ldots, \X_{m_d}(s) \right]^\top$. For this model, the estimate of the vector of regression coefficient functions is given by
\begin{equation*}
\widehat{\beta}^{\mathcal{M}}_{\tau}(s,t) = \underset{\begin{subarray}{c}
  \bm{\beta}^{\mathcal{M}}_{\tau}(s,t)
  \end{subarray}}{\argmin} \left[ \sum_{i=1}^n \rho_{\tau} \left( \Y_i(t) - \int_0^1 \left( \bm{\X}^{\mathcal{M}}_i(s)\right)^\top \bm{\beta}^{\mathcal{M}}_{\tau}(s,t) ds \right) \right]. 
\end{equation*}
Denote by $\vert \mathcal{M} \vert$ the cardinality $d$ of $\mathcal{M}$. Then, similar to \cite{Lee2014}, the BIC for this model can be defined
\begin{equation}\label{eq:bicvs}
\text{BIC} \left( \mathcal{M} \right) = \ln \bigg \Vert \left[ \sum_{i=1}^n \rho_{\tau} \left( \Y_i(t) - \int_0^1 \left( \bm{\X}^{\mathcal{M}}_i(s) \right)^\top \bm{\beta}^{\mathcal{M}}_{\tau}(s,t) ds \right) \right] \bigg \Vert_{\mathcal{L}_2} + \vert \mathcal{M} \vert \frac{\ln (n)}{2 n}.
\end{equation}
Based on the BIC in~\eqref{eq:bicvs}, we consider the following forward stepwise procedure to determine significant functional predictors:

\begin{itemize}
\item[1)] Construct $M$-function-on-function linear QR model models using the common response and a functional predictor:
\begin{equation*}
Q_{\tau}\left[ \Y_i(t) \big\vert \bm{\X}_i(s) \right] = \int_0^1 \left( \bm{\X}^{\mathcal{M}}_i(s)\right)^\top \bm{\beta}^{\mathcal{M}}_{\tau}(s,t) ds,
\end{equation*}
where $\bm{\X}^{\mathcal{M}}_i(s) = \X_{im}(s)$ for $m = 1, \ldots, M$. Then, we determine the model with the smallest $\text{BIC} \left( \mathcal{M} \right)$ value among these models as the initial model. Denote by $\bm{\X}^{\mathcal{M}_{(1)}}_i(s)$ and $\text{BIC}^{(1)}\left( \mathcal{M} \right)$ the predictor variable in the initial model and the BIC value calculated from this model, respectively.
\item[2)] Similar to Step 1), we construct $(M-1)$-function-on-function linear QR model as follows:
\begin{equation*}
Q_{\tau}\left[ \Y_i(t) \big\vert \bm{\X}_i(s) \right] = \int_0^1 \left( \bm{\X}^{\mathcal{M}}_i(s)\right)^\top \bm{\beta}^{\mathcal{M}}_{\tau}(s,t) ds,
\end{equation*}
where $\bm{\X}^{\mathcal{M}}_i(s) = \left[ \bm{\X}_{im}^{(1)}(s), \X_{im}(s) \right]$ and $\X_{im}(s) \neq \bm{\X}_{im}^{(1)}(s)$, and calculate the BIC values for each of these models. Let $\text{BIC}^{(2)}\left( \mathcal{M} \right)$ denote the smallest BIC value calculated from these $(M-1)$ models. Then, the predictor vector, $\bm{\X}^{\mathcal{M}_{(2)}}_i(s)$ corresponding to $\text{BIC}^{(2)}\left( \mathcal{M} \right)$ is chosen as the predictor vector for the current model if $\text{BIC}^{(2)}\left( \mathcal{M} \right) / \text{BIC}^{(1)}\left( \mathcal{M} \right) < 0.9$. In other words, the second functional predictor is included to the model if it contributes at least 10\% to the model. We determine the threshold value ``10\%'' based on the Monte Carlo experiments performed in this study (our results show that the variable selection procedure generally determines the all the significant functional predictors when the threshold value is 10\%).
\item[3)] Step 2) is repeated until all the significant functional predictors are determined.
\end{itemize}

Note that the joint selection of the variable selection and determination of the optimum number of FPQR components may not be computationally efficient when a large number of predictors are considered. In our numerical analyses, we first determine the significant predictor variables using only the first FPQR component. Conditional on the significant predictors, we then select the number of retained components. Our numerical results suggest that the choice of a fixed number of components does not significantly affect the determination of significant predictor variables. 

\section{Numerical results} \label{sec:results}

\subsection{Monte Carlo simulations}\label{sec:mc}

\begin{figure}[!htb]
  \centering
  \includegraphics[width=8.9cm]{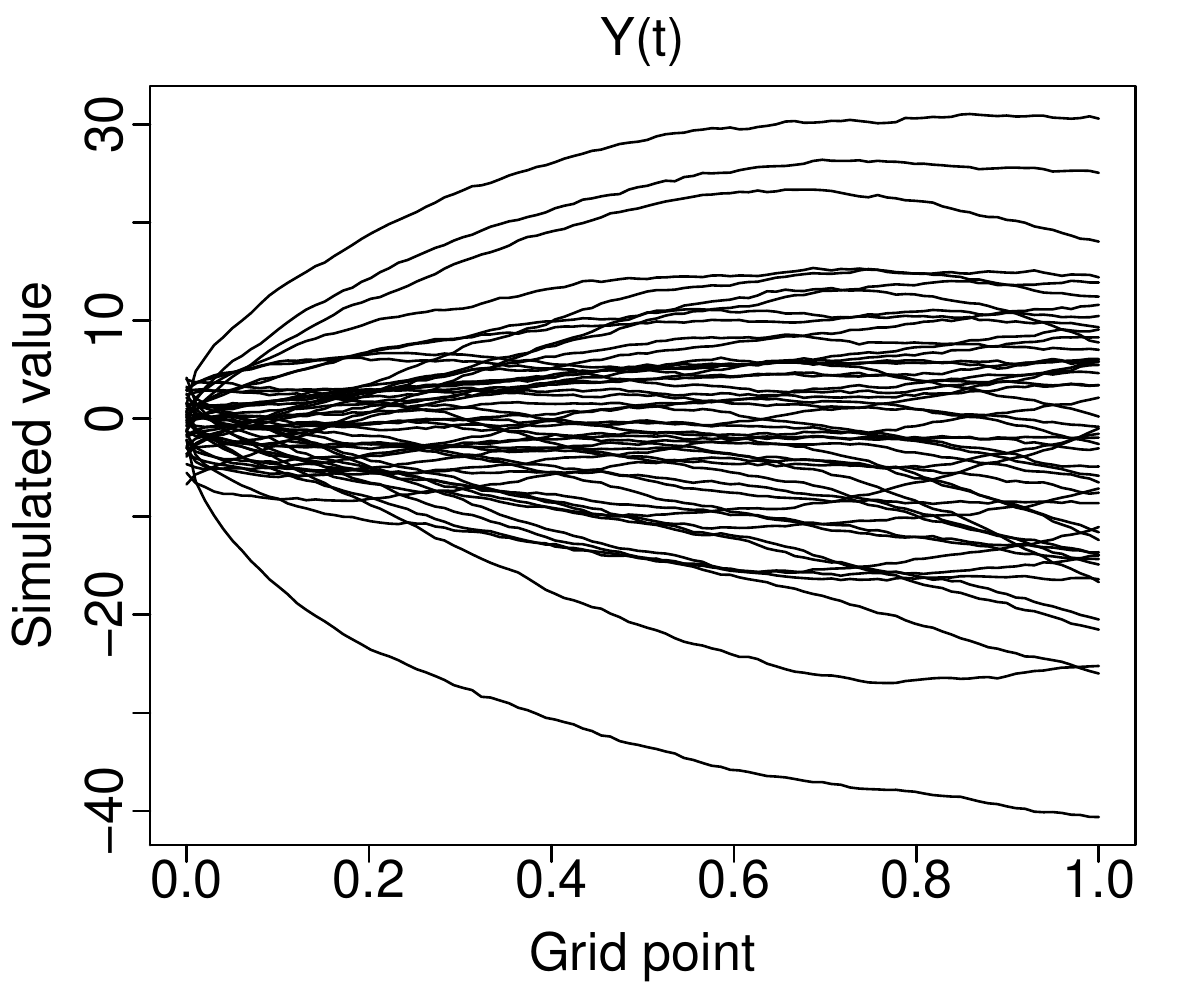}
  \includegraphics[width=8.9cm]{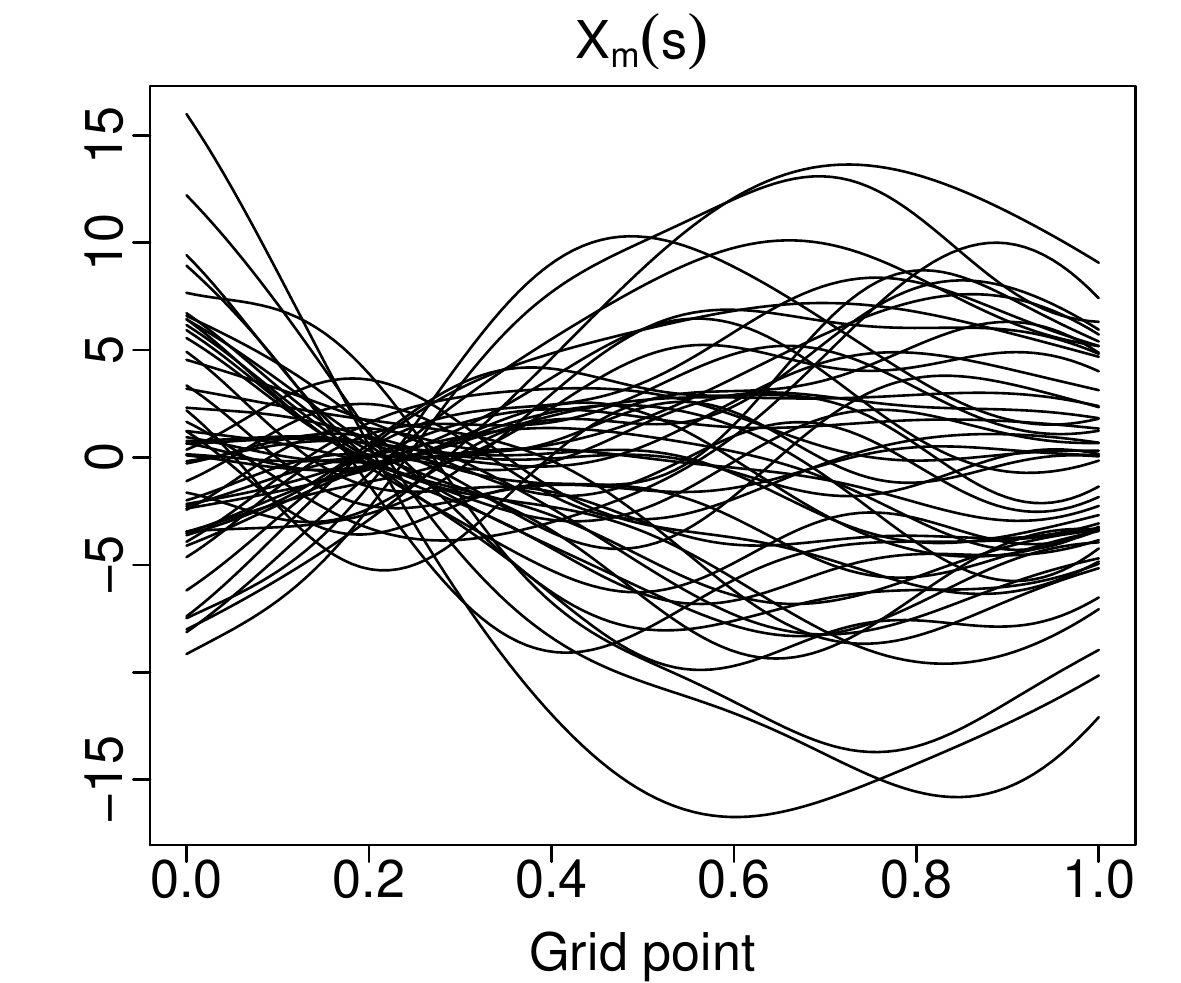}
  \caption{Graphical display of the generated 50 functions of the functional response (left panel) and functional predictor (right panel).}
  \label{fig:Fig_1}
\end{figure}

Several Monte Carlo simulations are performed to assess the finite-sample performance of the proposed FPQR method, and the results are compared with those obtained via the FPLS \citep{BSenv}. Throughout the experiments, MC = 200 Monte Carlo simulation runs are performed, and for each run, $M = 5$ functional predictors with sample sizes $n = [400, 550, 800]$ are generated at 100 equally spaced points in the interval $[0, 1]$. The following process is used to generate the functional predictors:
\begin{equation*}
\X_m(s) = \sum_{i=1}^5 \xi_i \varphi_i(s),
\end{equation*}
where $\xi_i \sim N\left( 0, 4 i^{-\frac{3}{2}} \right)$ and $\varphi_i(s) = \sin(i \phi s) - \cos(i \phi s)$. The smooth bivariate coefficient functions $\beta_m(s,t)$ are generated as follows:
\begin{align*}
\beta_1(s,t) &= 2 \sin( 2 \pi s) \sin (\pi t), \\
\beta_2(s,t) &= \cos \left( \frac{3}{2} \pi s \right) \cos \left(\frac{3}{2} \pi t \right),  \\
\beta_3(s,t) &= e^{-(s - 1/2)^2} e^{-2 (t-1)^2} \\
\beta_4(s,t) &= (s - 1/2)^2 (t - 1/2)^2 \\
\beta_5(s,t) &= 4 \sqrt{s} \sqrt{2 t}.
\end{align*}
Then, the functional response is generated as the linear combinations of the functional predictors and smooth bivariate coefficient functions as follows:
\begin{equation*}
\Y(t) = \sum_{m \in \lbrace 1,2,5\rbrace} \int_0^1 \X_m(s) \beta_m(s,t) + \epsilon(t),
\end{equation*}
where the error term $\epsilon(t)$ is generated using the Ornstein-Uhlenbeck process:
\begin{equation*}
\epsilon(t) = \gamma + [\epsilon_0(t) - \gamma] e^{-\theta t} + \sigma \int_0^t e^{-\theta(t-u)} d W_u,
\end{equation*}
where $\gamma \in \mathbb{R}$, $\theta >0$, and $\sigma > 0$ are real constants, and $W_u$ is the Wiener process. Herein, $\epsilon_0(t)$ denotes the initial value of $\epsilon(t)$ and is taken independently from $W_u$. The sample error functions are then obtained via sampling the joint distribution of $\epsilon(t)$. An example of the generated random functions is presented in Figure~\ref{fig:Fig_1}.

Throughout the simulations, the finite-sample performance of the proposed method are compared with the FPLS under three cases:
\begin{description}
\item[Case-1:] The functional response and functional predictors are generated as given above. In this case, the functional random variables are generated using a smooth data generation process with a normally distributed error term. In this case, the aim is to show if the proposed FPQR performs similarly to the FPLS.
\item[Case-2:] The functional predictors are generated as in Case-1, but the functional response is generated using $\chi^2_{(1)}$ distributed error term. To this end, the Ornstein-Uhlenbeck process is generated via sampling the joint $\chi^2_{(1)}$ distribution. In this case, the aim is to show if the proposed method outperforms the FPLS when the error term follows a non-Gaussian heavy-tailed distribution.
\item[Case-3:] The magnitude outliers contaminate 10\% of the functions of the generated data. Since the QR focuses mainly on the characterization of the response variable's conditional distribution, only the functional response variable's observations are contaminated by the magnitude outliers. We consider a contaminated Ornstein-Uhlenbeck process to generate magnitude outliers. While doing so, $n\times$ 90\% of the error functions are generated as in Case-1, but the remaining $n\times$ 10\% (randomly selected) error functions are generated when the mean of initial value $\epsilon_0(t)$ is five and the similar standard deviation as in Case-1. In this case, the aim is to show if the proposed method outperforms the FPLS when outliers are present in the data.
\end{description}

The predictive performance of the FPQR and FPLS are compared using the mean squared prediction error (MSPE) measure. To this end, the following procedure is considered. 
\begin{enumerate}
\item[1)] We divide the entire generated data into training and test samples. To evaluate the effect of sample size on the finite-sample performance of the methods, the size of the test sample ($n_{\text{test}}$) is fixed at 300 and three different sizes are considered for the training sample; $n_{\text{train}} = [100, 250, 500]$.
\item[2)] We estimate the $\tau\textsuperscript{th}$ conditional quantile of the functional response given the functional predictors using the functions in the training sample.
\item[3)] Then, we calculate the MSPE values based on the estimated conditional quantile of the functional response and the test sample's functional predictors as follows:
\begin{equation*}
\text{MSPE} = \frac{1}{n_{\text{test}}} \sum_{i=1}^{n_{\text{test}}} \Vert \Y_i(t) - \widehat{\Y}_i(t) \Vert_{\mathcal{L}_2}^2,
\end{equation*}
where $\widehat{\Y}_i(t)$ is the predicted functional response for $i^\textsuperscript{th}$ individual.
\end{enumerate}

Note that, for both the FPQR and FPLS, the MSPE values are calculated under three models:
\begin{inparaenum}
\item[1)] The \textit{full model} where all the five generated functional predictors are used to estimate $Q_{\tau}\left[ \Y_i(t) \big\vert \bm{\X}_i(s) \right]$;
\item[2)] the \textit{true model} where only the significiant functional predictors $\left\lbrace \X_1(s), \X_2(s), \X_5(s) \right\rbrace$ are used to estimate $Q_{\tau}\left[ \Y_i(t) \big\vert \bm{\X}_i(s) \right]$; and
\item[3)] the \textit{selected model} where the significiant functional variables determined by the variable selection method introduced in Section~\ref{sec:vs} are used to estimate $Q_{\tau}\left[ \Y_i(t) \big\vert \bm{\X}_i(s) \right]$.
\end{inparaenum}
Since the FPLS is a mean regression, the quantile level $\tau = 0.5$ is considered in the simulations, and the focus is restricted to evaluate functional mean regression versus functional median regression. In the simulations, $K = 10$ numbers of basis functions for all functional variables are used to project them into finite-dimensional space.

To investigate the forecast uncertainties obtained by both methods, the following case-sampling-based bootstrap approach is used to construct pointwise prediction intervals.
\begin{itemize}
\item[Step 1)] We obtain a bootstrap sample $\left[ \Y^*(t), \bm{\X}^*(s)\right]$ by sampling with replacement from the pair $\left[ \Y(t), \bm{\X}(s)\right]$.
\item[Step 2)] We estimate the $\tau\textsuperscript{th}$ conditional quantile of the functional response, $\widehat{Q}_{\tau}^*\left[ \Y_i(t) \big\vert \bm{\X}_i(s) \right]$ using the bootstrap sample.	
\item[Step 3)] We obtain $B$ bootstrap replicates of $Q_{\tau}\left[ \Y_i(t) \big\vert \bm{\X}_i(s) \right]$, $ \left\lbrace \widehat{Q}_{\tau}^{*,b}\left[ \Y_i(t) \big\vert \bm{\X}_i(s) \right] \right\rbrace_{b=1}^B$, by repeating Steps 1-2 $B$ times.
\end{itemize}
Then, the $100(1-\alpha)\%$ bootstrap prediction interval for $Q_{\tau}\left[ \Y_i(t) \big\vert \bm{\X}_i(s) \right]$ is calculated as follows:
\begin{equation*}
\left[ Q_{\alpha/2}(t),~Q_{1-\alpha/2}(t) \right],
\end{equation*}
where $Q_{\alpha/2}(t)$ is the $\alpha/2\textsuperscript{th}$ quantile of the generated $B$ sets of bootstrap replicates. Two bootstrap error measures: the coverage probability deviance (CPD) and the interval score (score), is considered to evaluate the performance of the pointwise bootstrap prediction intervals:
\begin{align*}
\text{CPD} &= (1 - \alpha) - \frac{1}{n_{\text{test}}} \sum_{i=1}^{n_{\text{test}}} \mathbb{1} \left\{  Q_{\alpha/2}^i(t) \leq \Y_i(t) \leq Q_{1 - \alpha/2}^i(t) \right\}, \\
\text{score} &= \frac{1}{n_{\text{test}}} \sum_{i=1}^{n_{\text{test}}} \left[ \left\{ Q_{1 - \alpha/2}^i(t) - Q_{\alpha/2}^i(t) \right\} \right. + \frac{2}{\alpha} \left( Q_{\alpha/2}^i(t) - \Y_i(t) \right) \mathbb{1} \left\{ \Y_i(t) < Q_{\alpha/2}^i(t) \right\} \\
&\hspace{2.7in}{+ \left. \frac{2}{\alpha} \left( \Y_i(t) - Q_{1 - \alpha/2}^i(t) \right) \mathbb{1} \left\{ \Y_i(t) > Q_{1 - \alpha/2}^i(t) \right\} \right]}.
\end{align*}
Throughout the simulations, $\alpha$ is set to 0.05 to obtain 95\% bootstrap prediction intervals. Note that an example \texttt{R} code for the proposed method is provided in an online supplement file.

\begin{figure}[!htb]
  \centering
  \includegraphics[width=5.9cm]{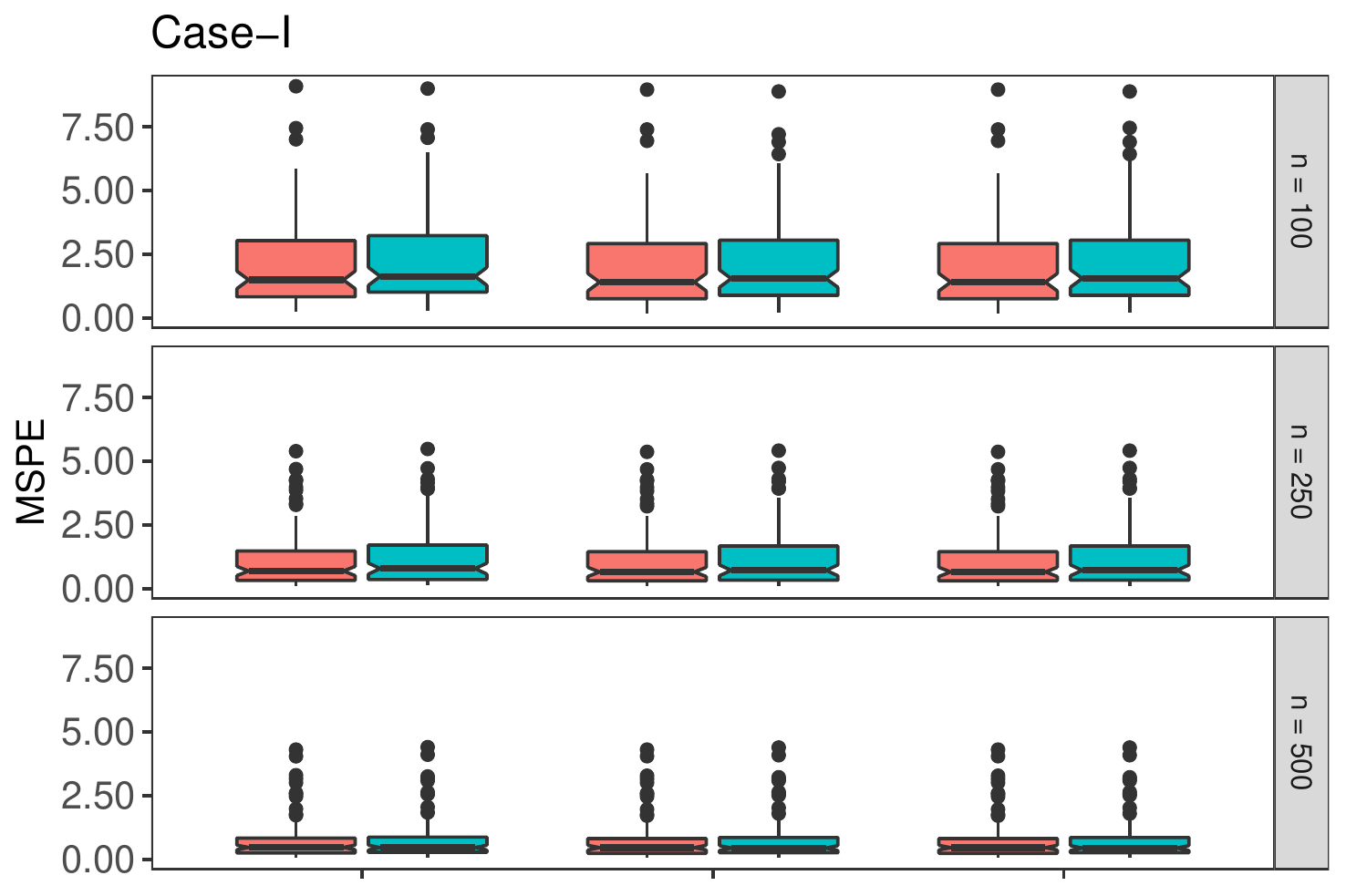}
  \includegraphics[width=5.9cm]{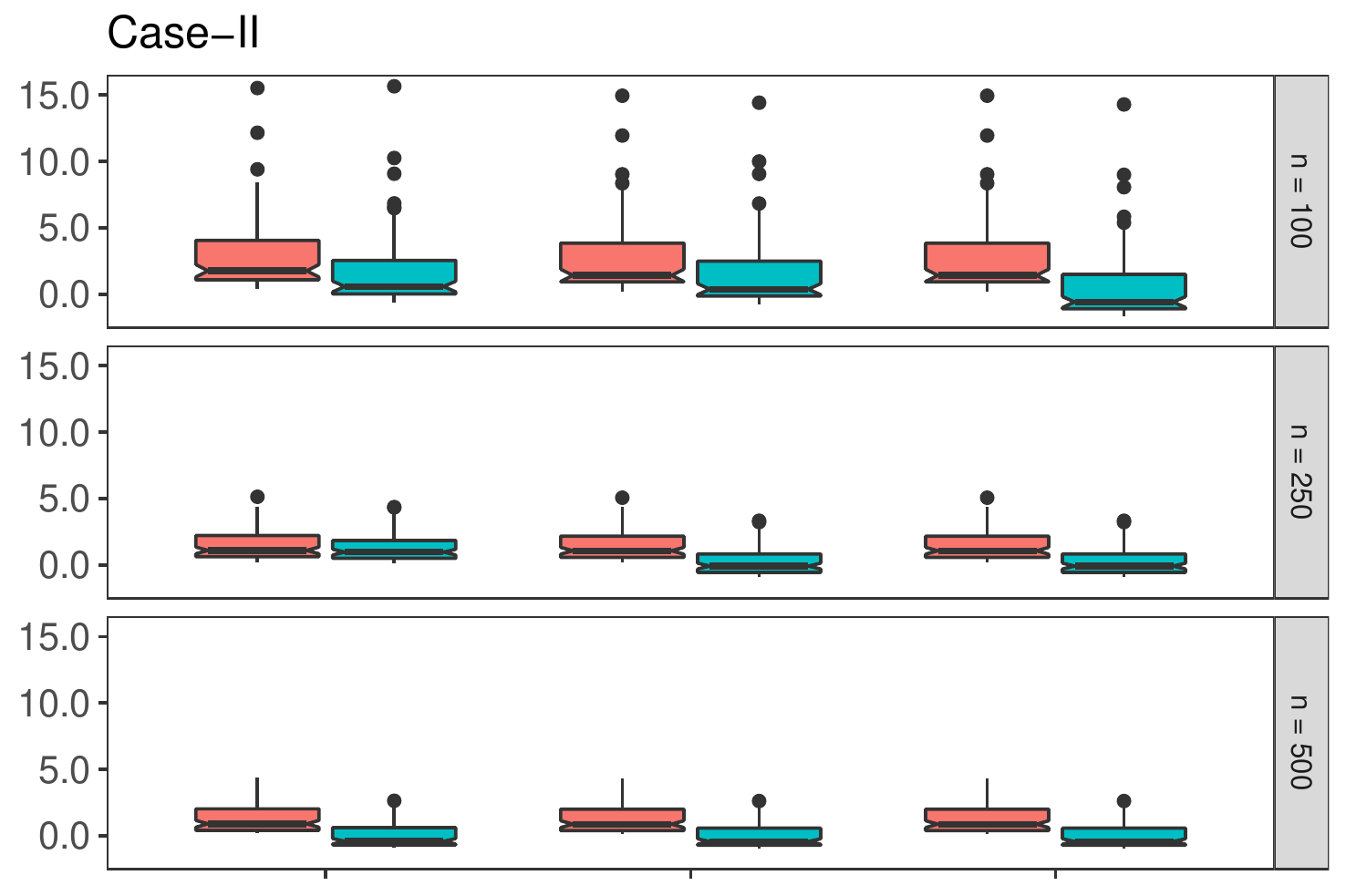}
  \includegraphics[width=5.9cm]{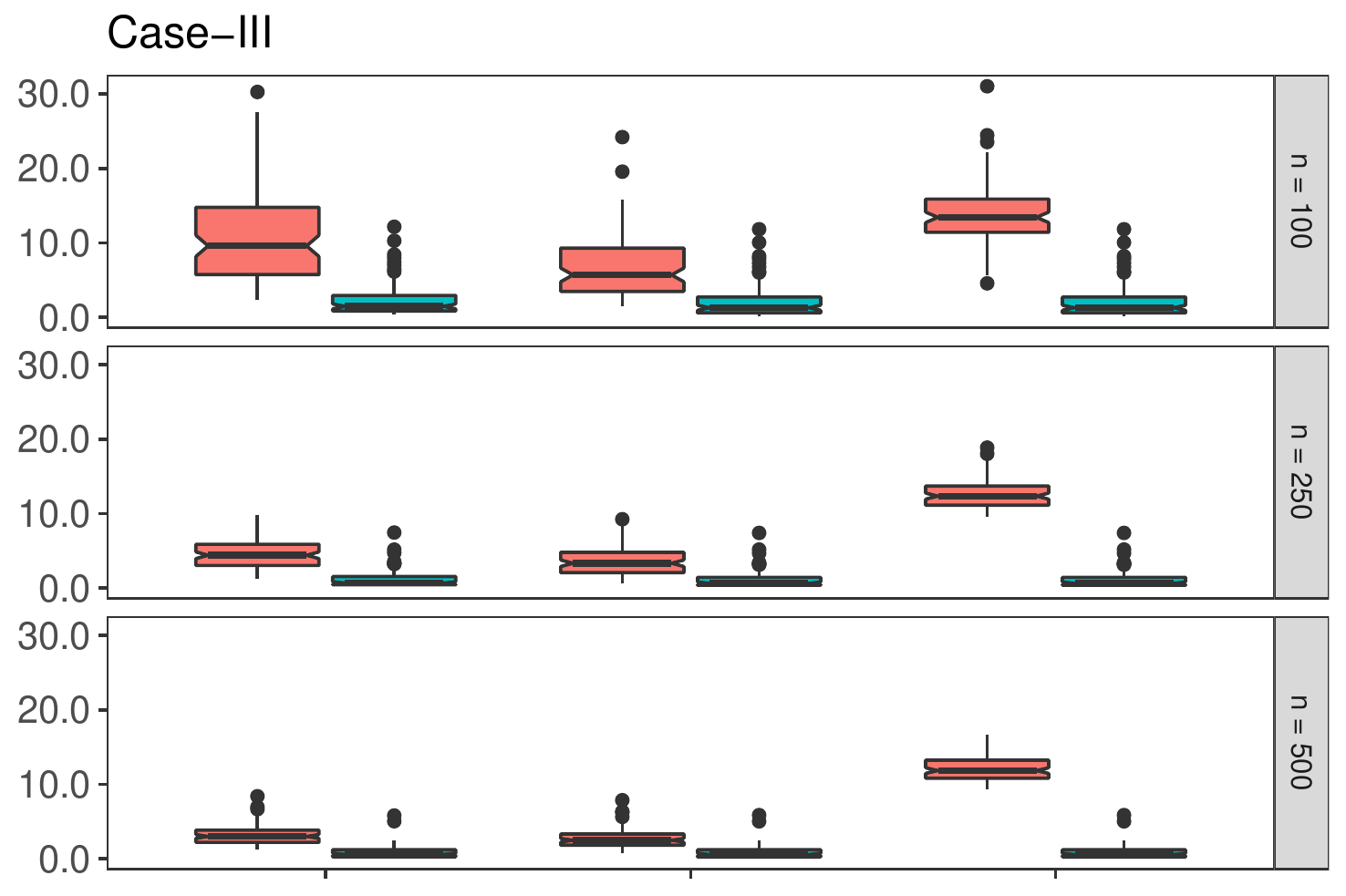}
\\  
  \includegraphics[width=5.9cm]{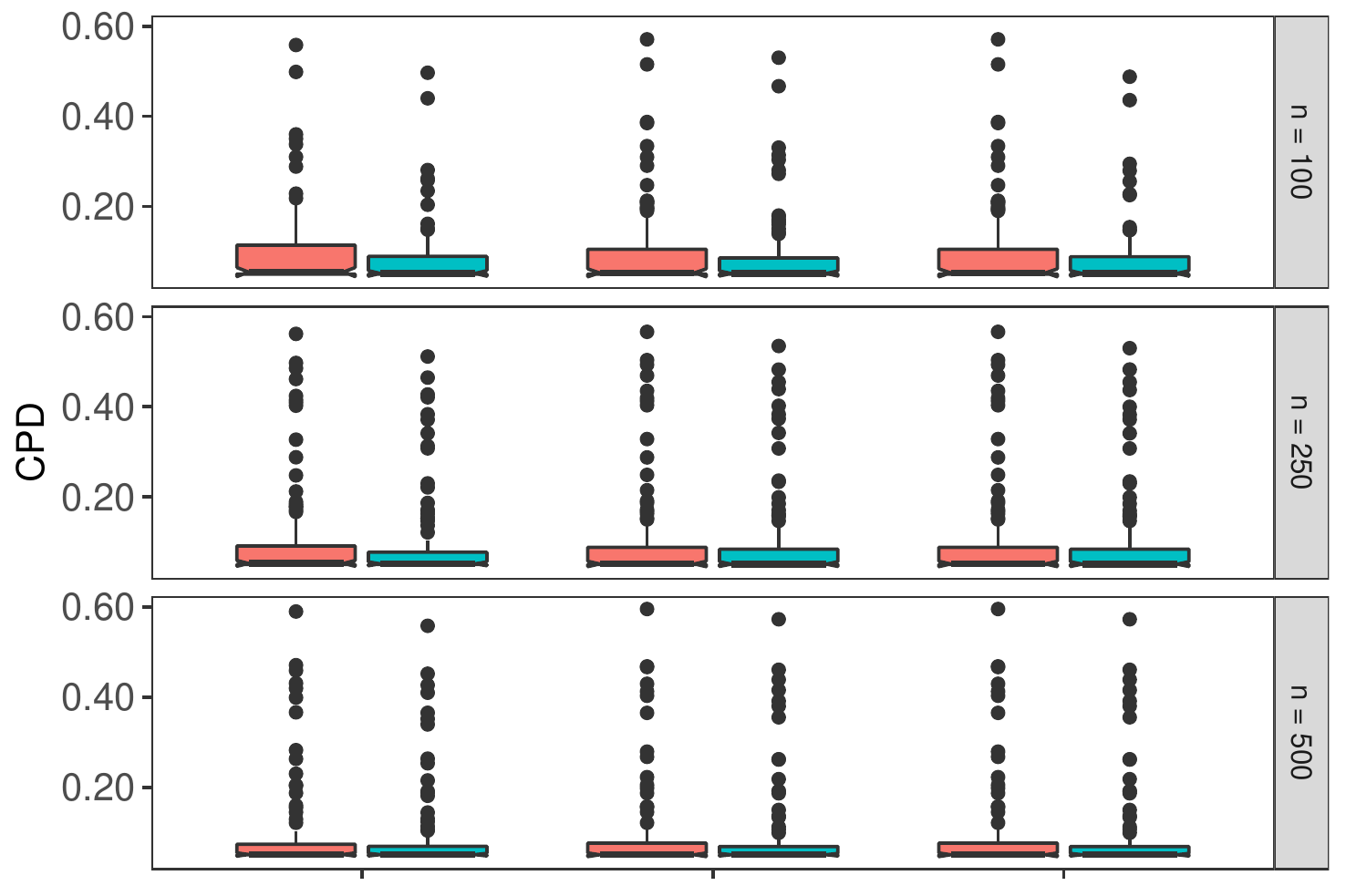}
  \includegraphics[width=5.9cm]{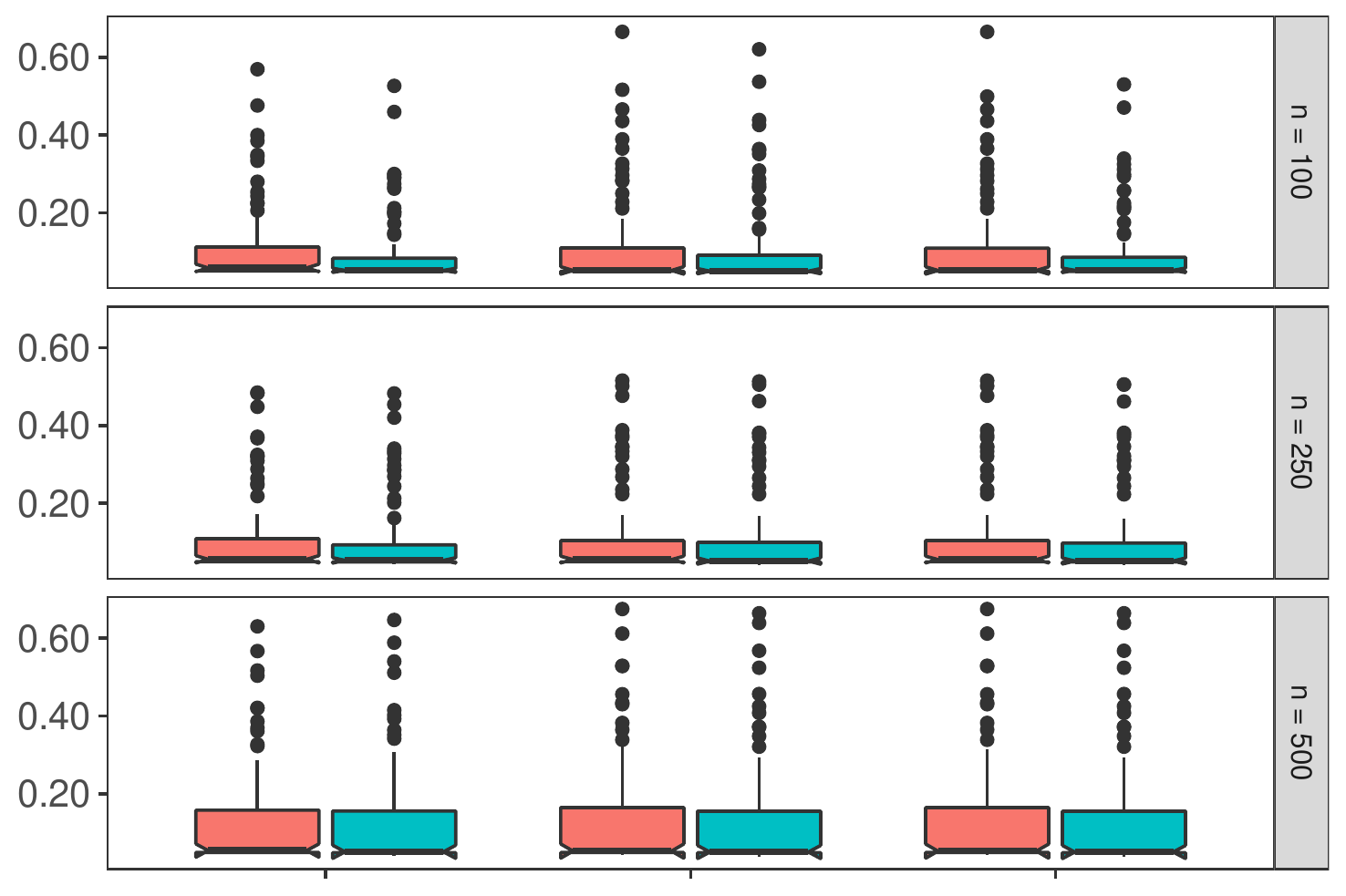}
  \includegraphics[width=5.9cm]{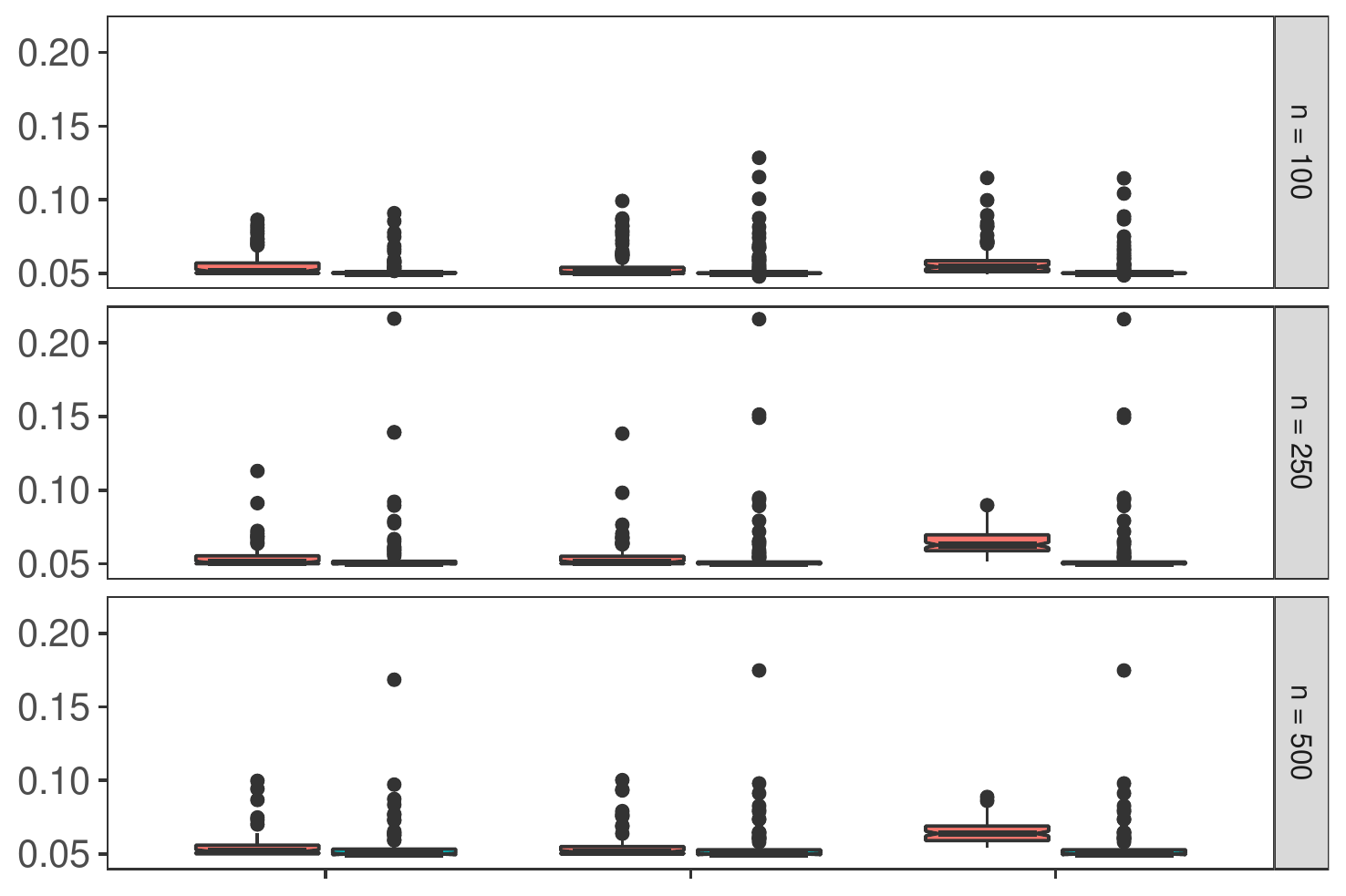}
\\  
  \includegraphics[width=5.9cm]{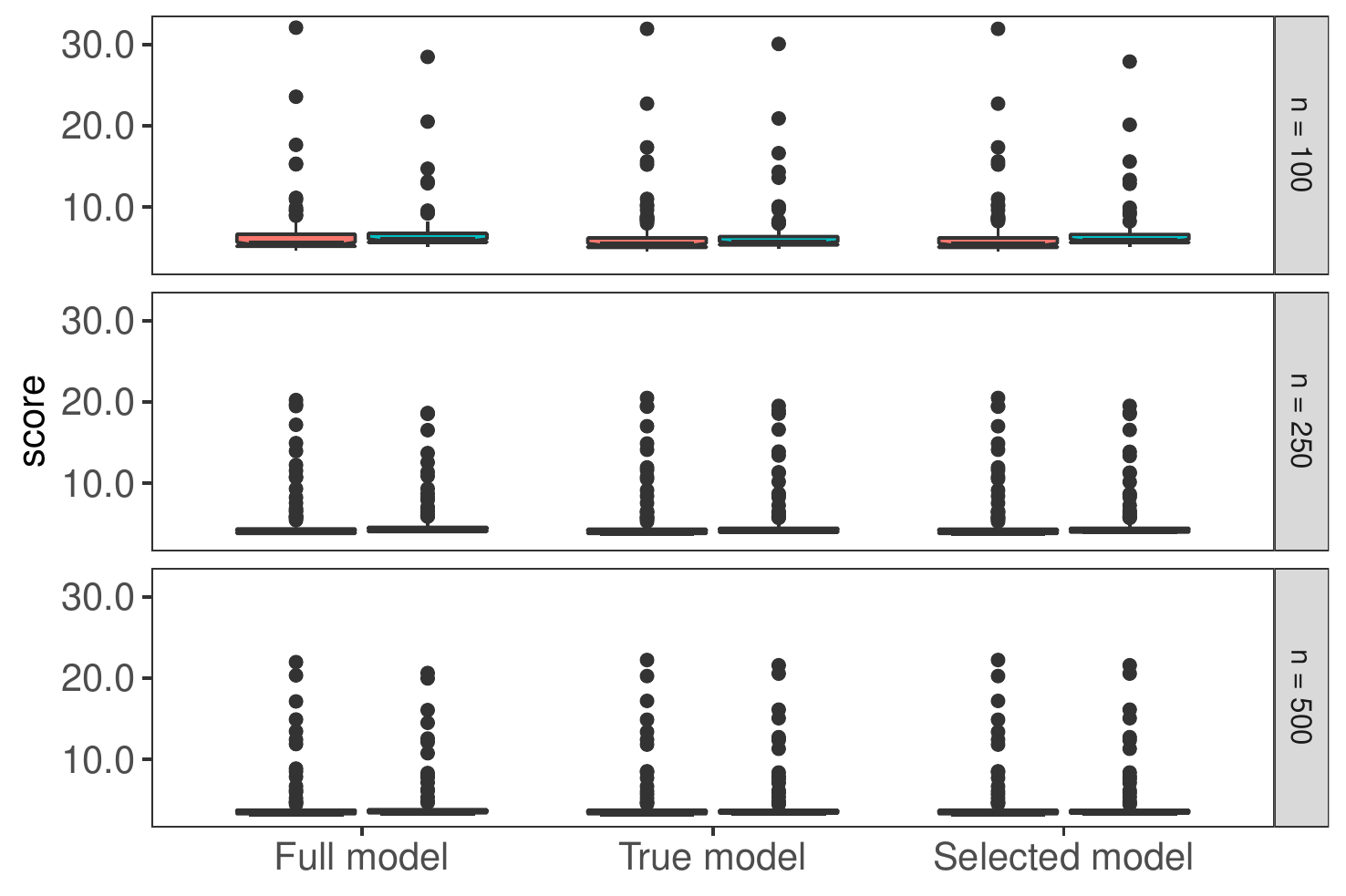}
  \includegraphics[width=5.9cm]{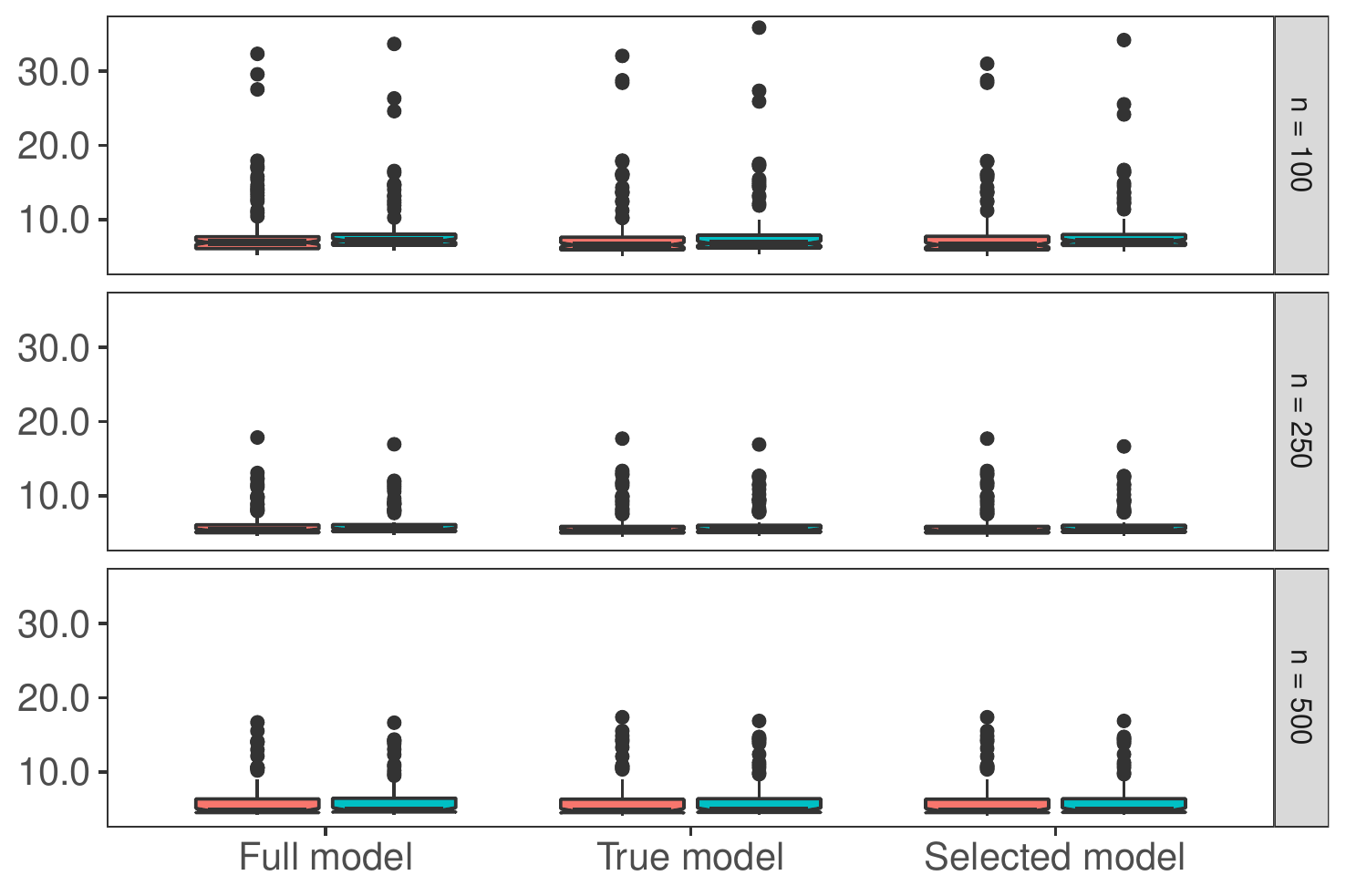}
  \includegraphics[width=5.9cm]{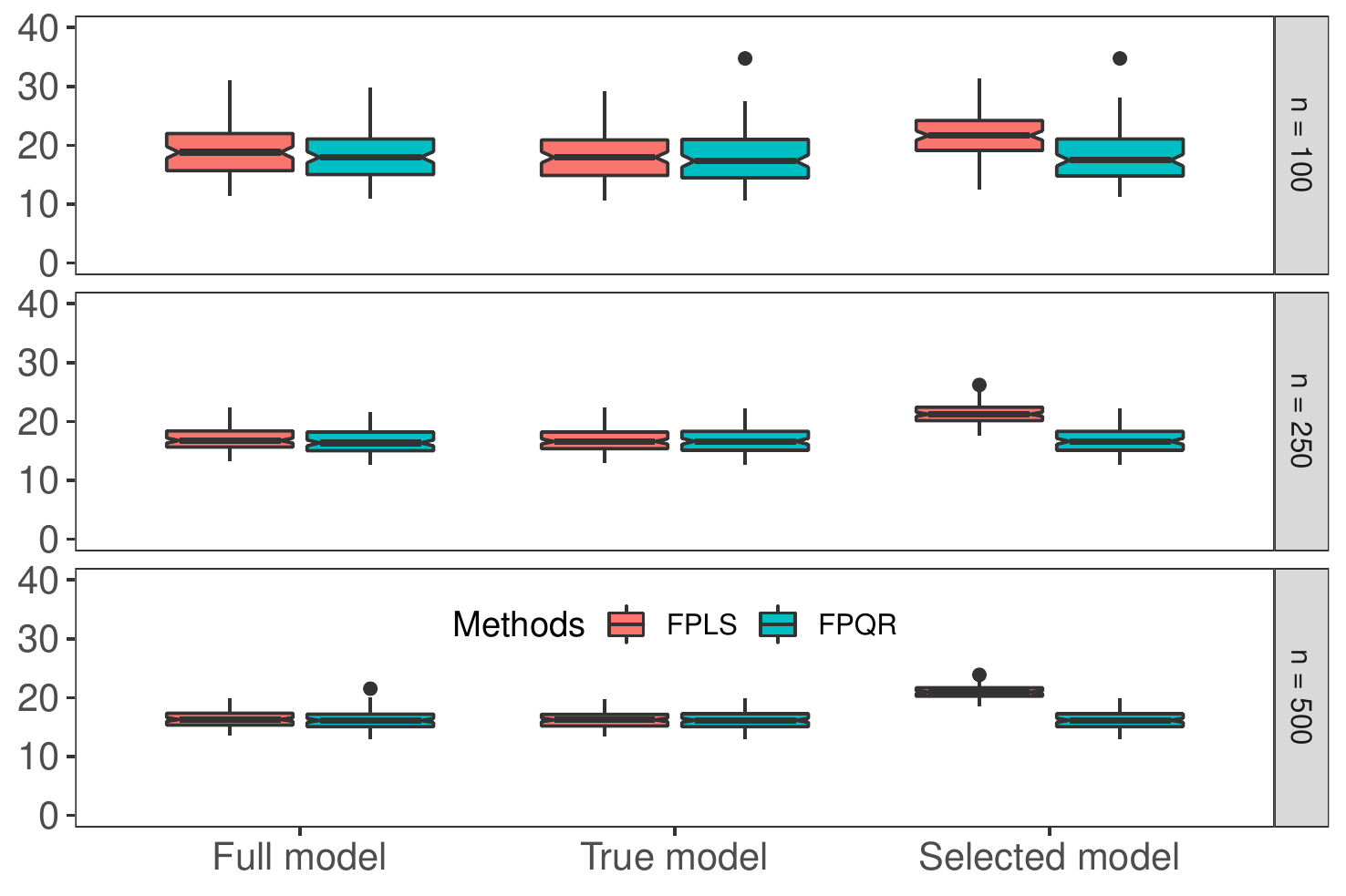}
  \caption{Predictive model performance: Computed MSPE (first row), CPD (second row), and score (third row) values of the FPLS and FPQR models under Case-1 (first column - Gaussian case), Case-2 (second column - heavy-tailed error case), and Case-3 (third column- outlier case).}
  \label{fig:Fig_2}
\end{figure}

The computed MSPE, CPD, and score values for all cases are presented in Figure~\ref{fig:Fig_2}. Our records indicate that, for all cases, the MSPE values computed under the actual and selected models are generally slightly smaller than those of the full model. Also, for both the FPLS and FPQR, the true and selected models produce almost the same MSPE values, which demonstrates that the variable selection method discussed in Section~\ref{sec:vs} has performed well in the determination of significant variables. When the error term follows the Gaussian distribution and no outliers are present in the data (Case-1), the proposed FPQR produces similar MSPE values with the FPLS. On the other hand, when the error terms follow $\chi^1_{(1)}$ distribution and no outliers are present in the data (Case-2), our proposed method produces smaller MSPE values than the FPLS under all the models. When the magnitude outliers are present in the data (Case-3), the results demonstrate that the proposed FPQR is robust to outliers, while the FPLS is significantly affected by these outliers. In this case, the FPQR produces considerably smaller MSPE values compared with the FPLS. In addition, Figure~\ref{fig:Fig_2} shows that both methods produce smaller MSPE values with the increasing sample sizes.

From Figure~\ref{fig:Fig_2}, the proposed method generally produces similar bootstrap-based CPD and interval score values with the FPLS. However, when magnitude outliers contaminate the data, the proposed method produces smaller CPD values with smaller score values than those of FPLS under the selected model. In other words, this result demonstrates that, compared with FPLS, the FPQR produces more accurate pointwise prediction intervals for the conditional quantiles of the response function with narrower prediction interval lengths when outliers are present in the data. All in all, the results produced by the Monte Carlo experiments performed in this study demonstrate that the proposed method produces similar performance with the FPLS when the errors follow a Gaussian distribution and no outliers are present in the data. However, it outperforms the FPLS when the errors follow a non-Gaussian heavy-tailed distribution or outliers are included in the data.

\subsection{Air quality data}

\begin{figure}[!htb]
  \centering
  \includegraphics[width=4.43cm]{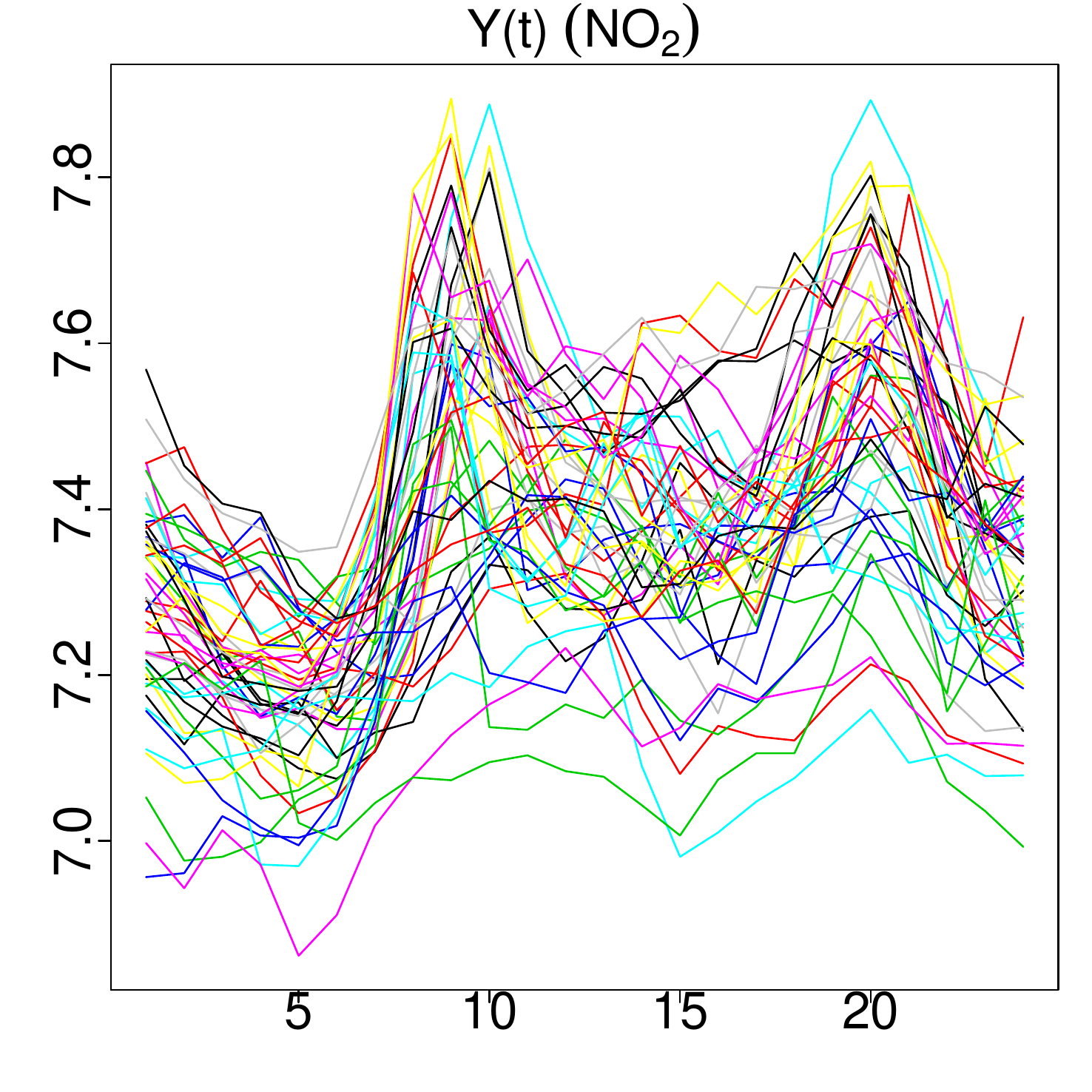}
  \includegraphics[width=4.43cm]{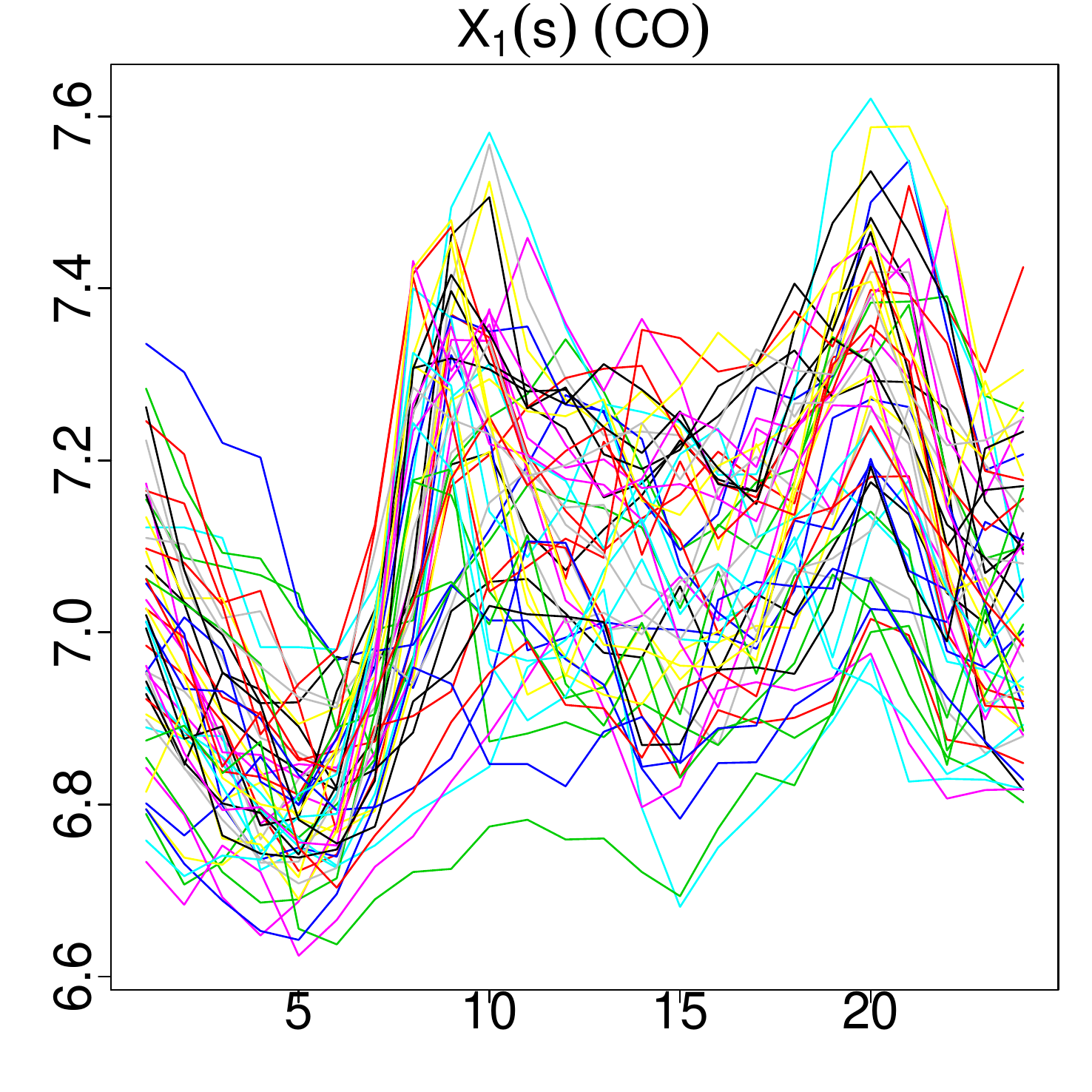}
  \includegraphics[width=4.43cm]{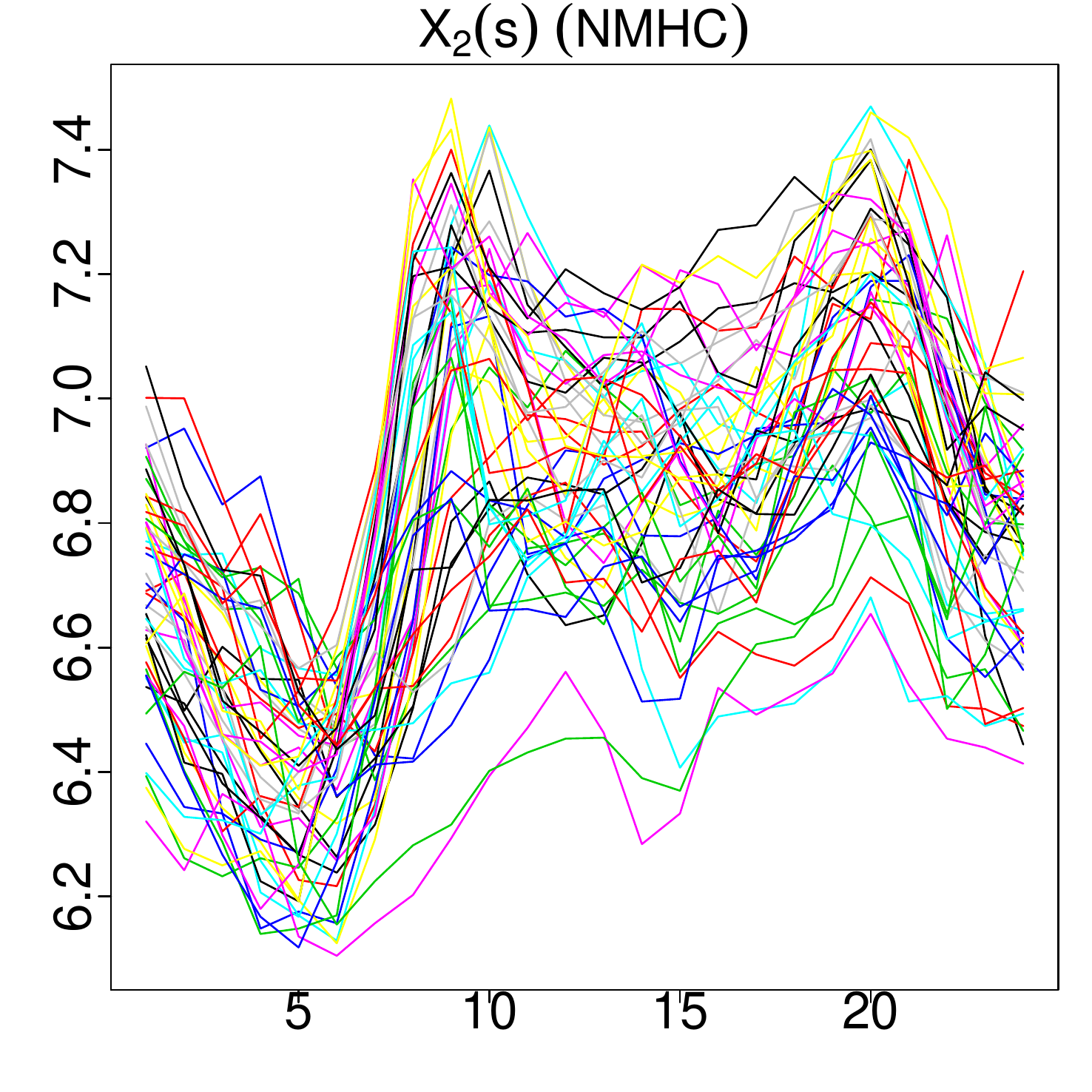}
  \includegraphics[width=4.43cm]{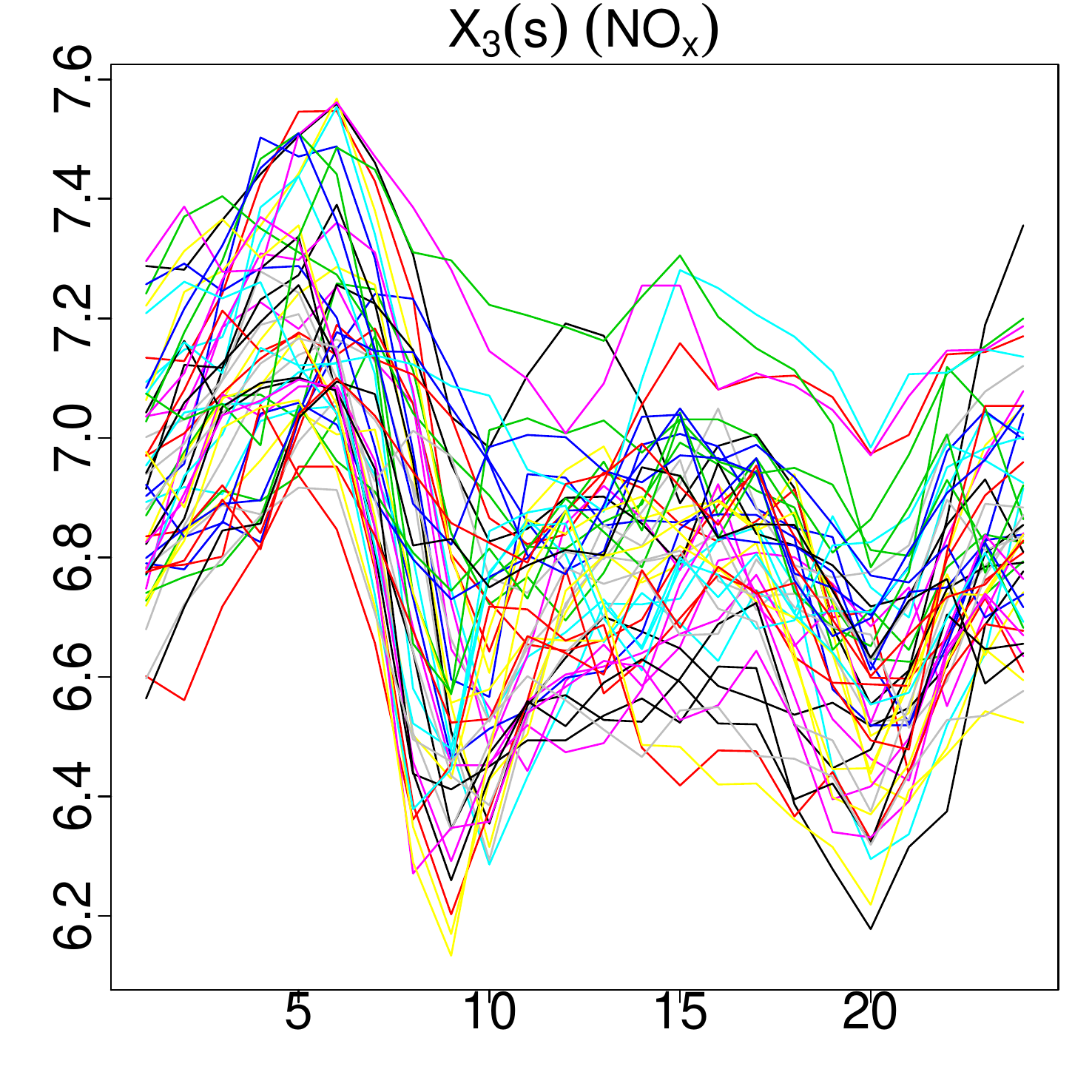}
\\  
  \includegraphics[width=4.43cm]{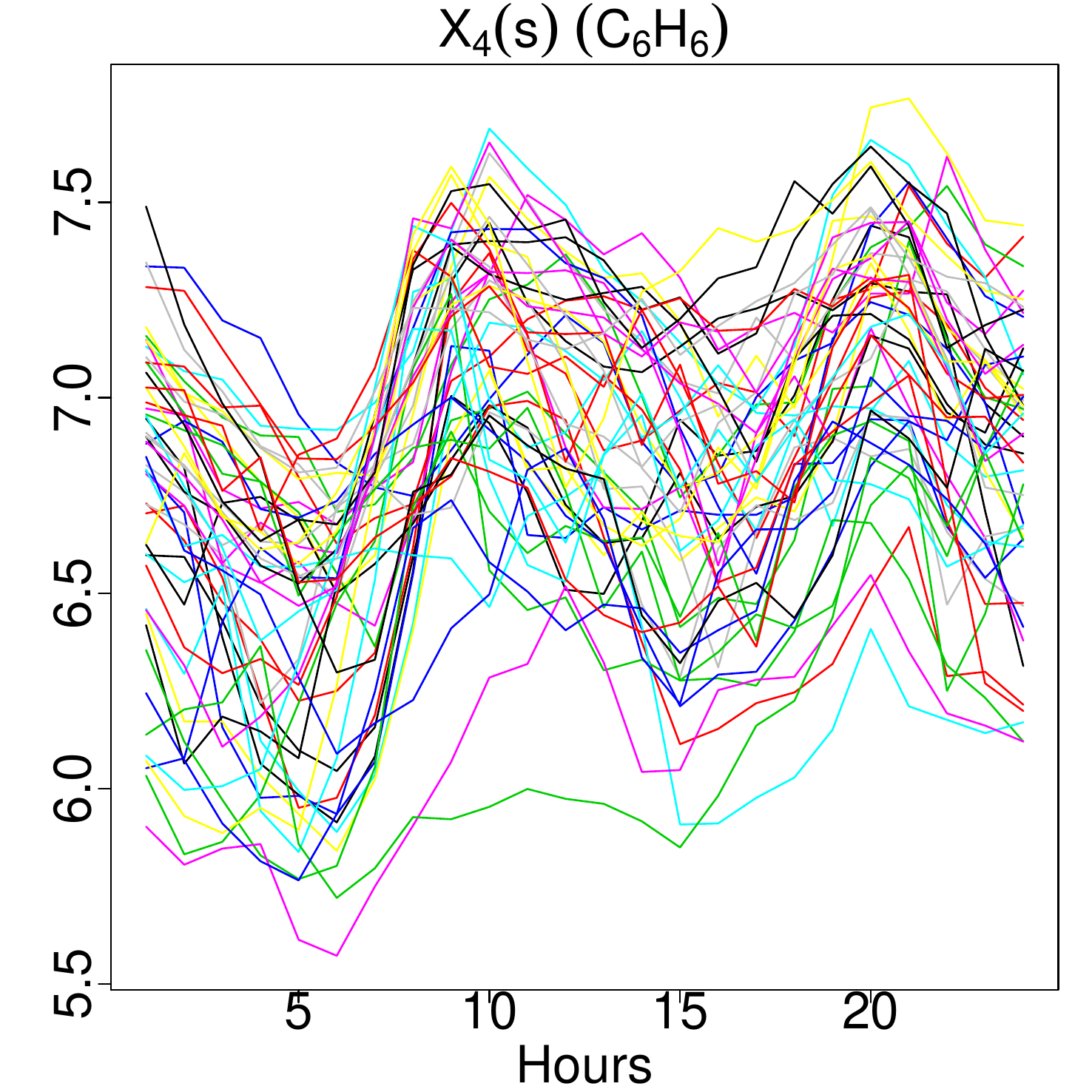}
  \includegraphics[width=4.43cm]{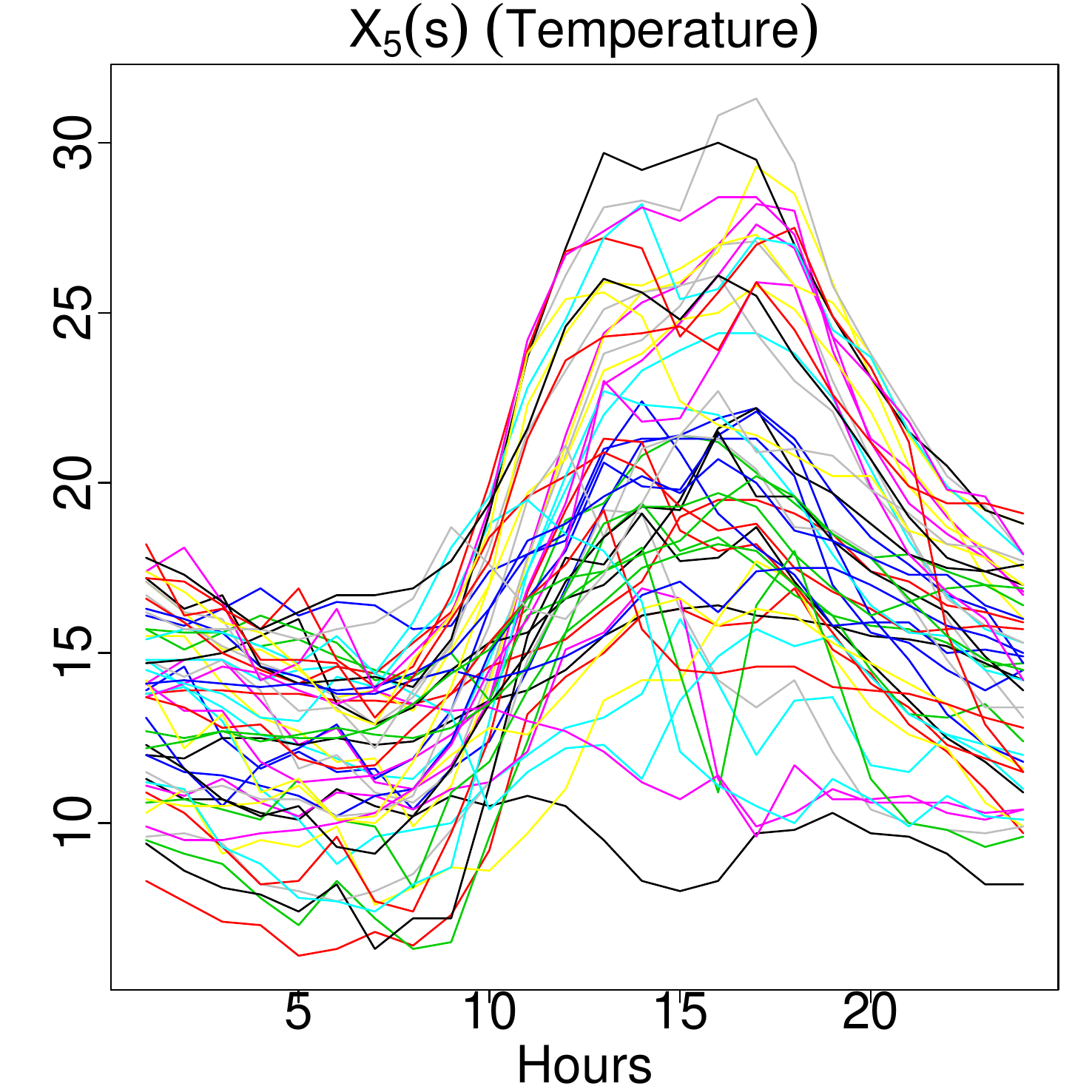}
  \includegraphics[width=4.43cm]{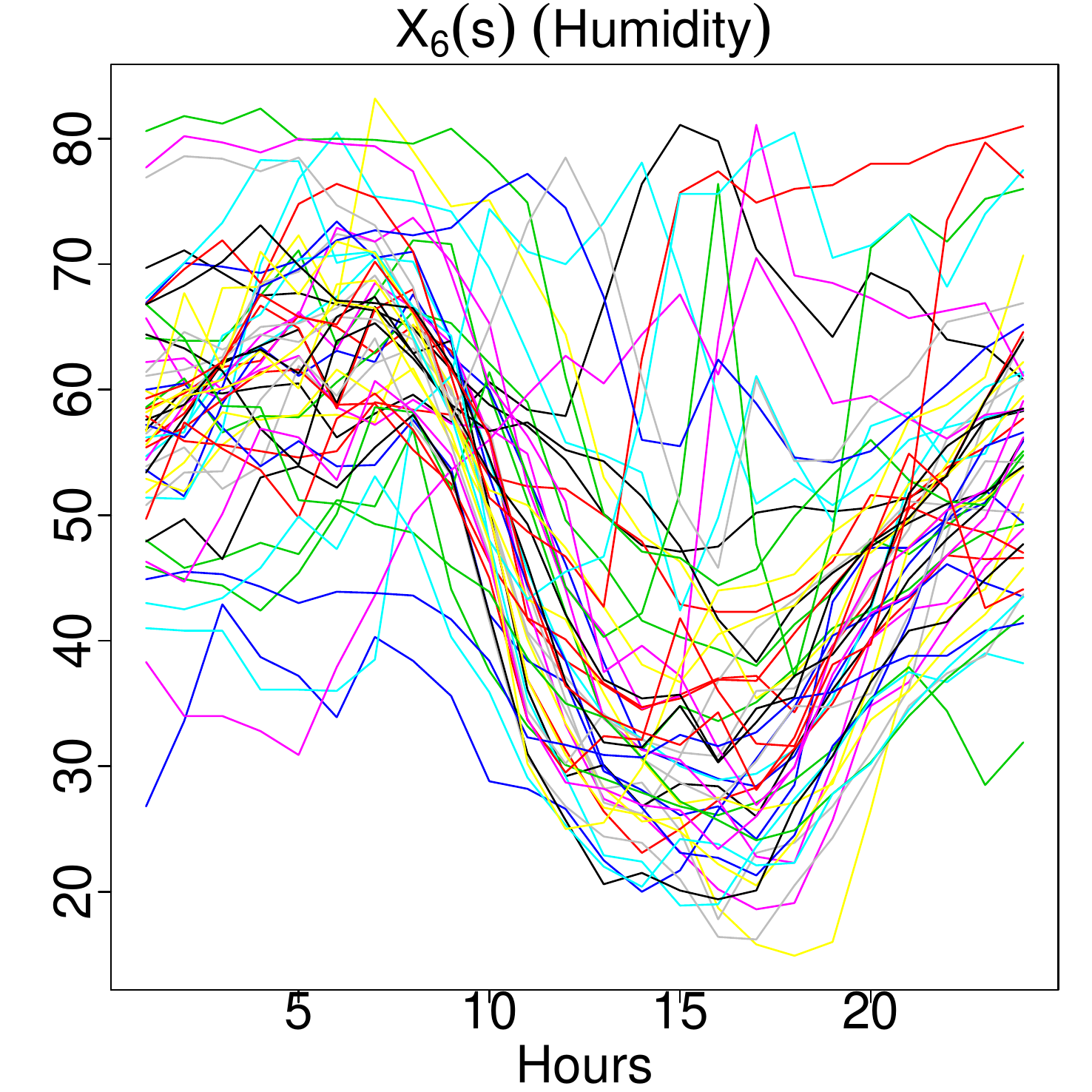}
  \caption{Graphical display of the functional variables; nitrogen dioxide (NO$_2$), carbon monoxide (CO), non-methane hydrocarbons (NMHC), total nitrogen oxides (NO$_x$), benzene (C$_6$H$_6$), temperature, and humidity. The observations are the functions of hours; $1 \leq s,t \leq 24$. Note that only the first 50 curves are presented for each variable, and different colors correspond to different hours.}
\label{fig:Fig_3}
\end{figure}

We consider an air quality dataset collected at the road level of a considerable pollutant Italian city \citep{Devito}. The dataset consists of hourly average concentration data for five different atmospheric pollutants each day. The hourly average concentration values were recorded by a multi-sensor device equipped with five metal oxide chemo resistive sensors. The atmospheric pollutants are the NO$_2$ (micro g/m$^3$), carbon monoxide (CO (mg/m$^3$)), non-methane hydrocarbons (NMHC (micro g/m$^3$)), total nitrogen oxides (NO$_x$ (ppb)), and benzene (C$_6$H$_6$ (micro g/m$^3$)). The dataset, which is available in the \texttt{R} package ``FRegSigCom'' \citep{LuoQi_R}, also includes the hourly temperature ($^o$C) and humidity (\%) values which were collected for each day. Each variable in this dataset consists of 355 curves observed at 24 equally spaced discrete time points in the interval $[1, 24]$. The graphical display of the six functional predictors and one functional response is presented in Figure~\ref{fig:Fig_3}.

We focus on investigating the functional relationship between NO$_2$ and other atmospheric pollutants and hydro climatological variables with the air quality dataset. In other words, we focus on predicting NO$_2$ curves for the given CO, NMCH, NO$_x$, C$_6$H$_6$, temperature, and humidity variables. To this end, we consider the following function-on-function regression model:
\begin{equation*}
\Y(t) = \beta_0(t) + \int_{s=1}^{24} \bm{\X}(s) \bm{\beta}(s,t) ds,
\end{equation*}
where $\beta_0(t)$ denotes the intercept function, $\bm{\X}(s) = \left\lbrace \X_1(s), \ldots, \X_6(s) \right\rbrace$, and $s,t \in [1,24]$. However, from Figure~\ref{fig:Fig_3}, the functional response and several functional predictors include potential outlying curves, which may lead incorrect prediction of the NO$_2$ curves. Consequently, compared with the function mean regression, we consider the proposed FPQR method may produce robust prediction of the NO$_2$ curves ($\tau = 0.5$). 

The following procedure is repeated 1000 times to compare the predictive performance of the proposed FPQR with the FPLS.
\begin{inparaenum}
\item[1)] We randomly divide the entire dataset into a training sample of size 155 and a testing sample of size 200.
\item[2)] With the training sample, we construct a model using both the FPQR and FPLS.
\item[3)] We then predict the remaining NO$_2$ curves in the test sample based on the constructed models and given CO, NMCH, NO$_x$, C$_6$H$_6$, temperature, and humidity curves in the testing sample.
\end{inparaenum}
For each replication, the MSPE value is computed for both methods. The case-sampling-based bootstrap method discussed in Section~\ref{sec:mc} is applied to construct pointwise confidence intervals for the NO$_2$ curves in the test sample, and CPD and score values are calculated. For the air quality datasets, the performance metrics are calculated under two models. The \textit{full model}, where all the atmospheric pollutants and hydro climatological variables are included in the model. The \textit{selected model}, where only the significant functional variables specified by the variable selection procedure discussed in Section~\ref{sec:vs} are used to construct the model. Also, note that $K = 16$ numbers of $B$-spline basis functions are used to construct their functional forms for all the functional variables.

\begin{figure}[!htb]
  \centering
  \includegraphics[width=7.6cm]{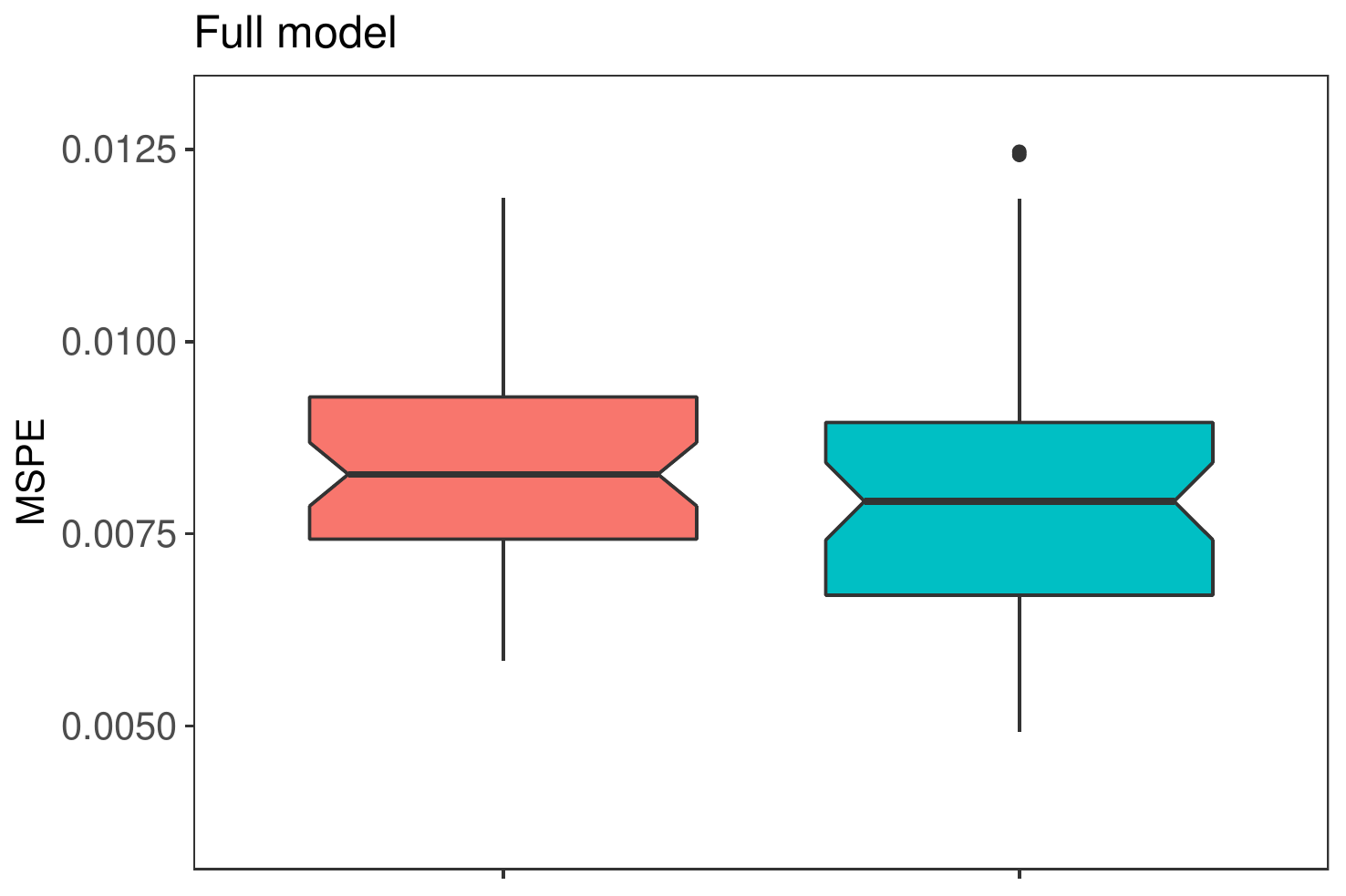}
\qquad
  \includegraphics[width=7.6cm]{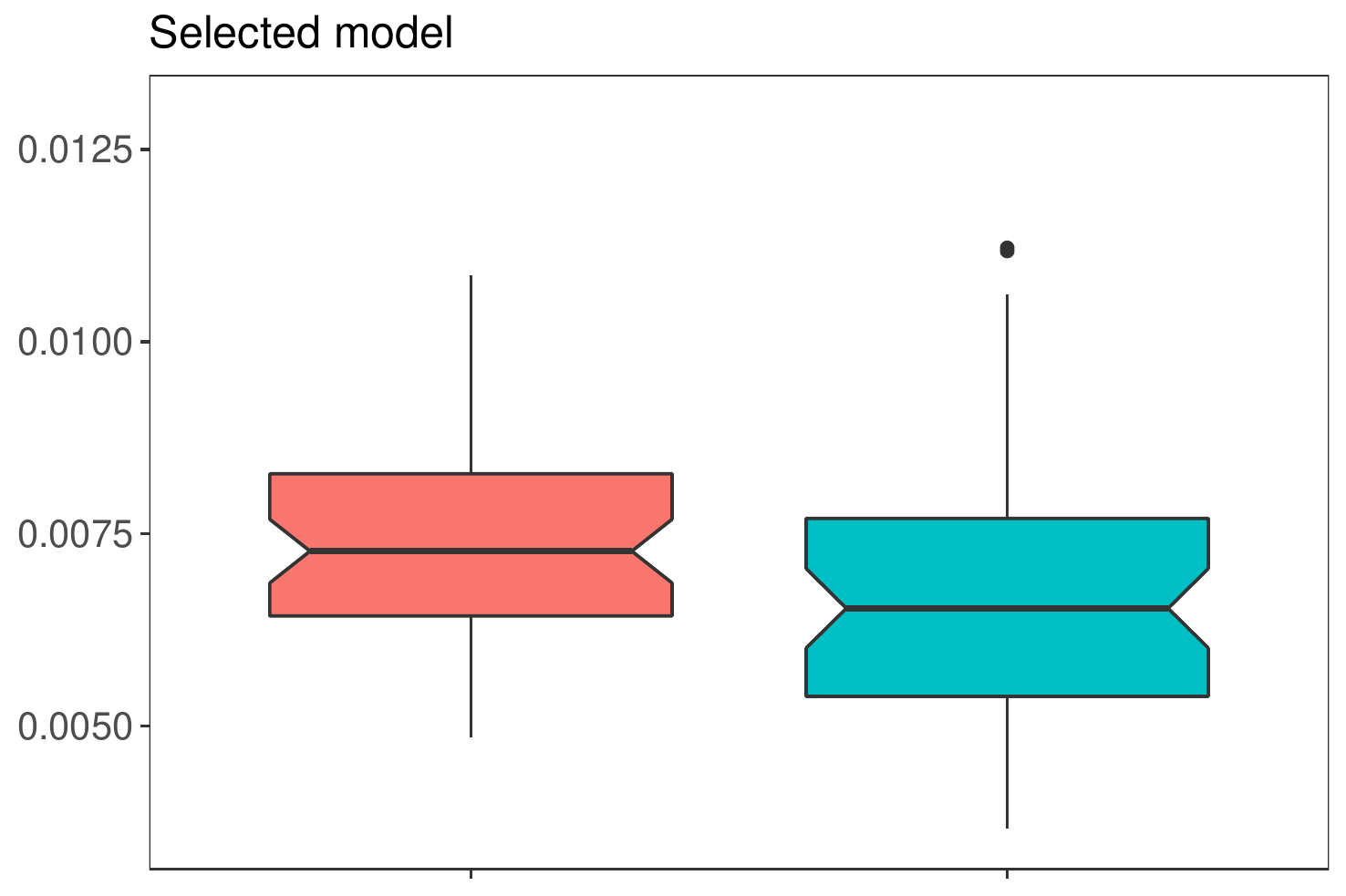}
  \\
  \includegraphics[width=7.6cm]{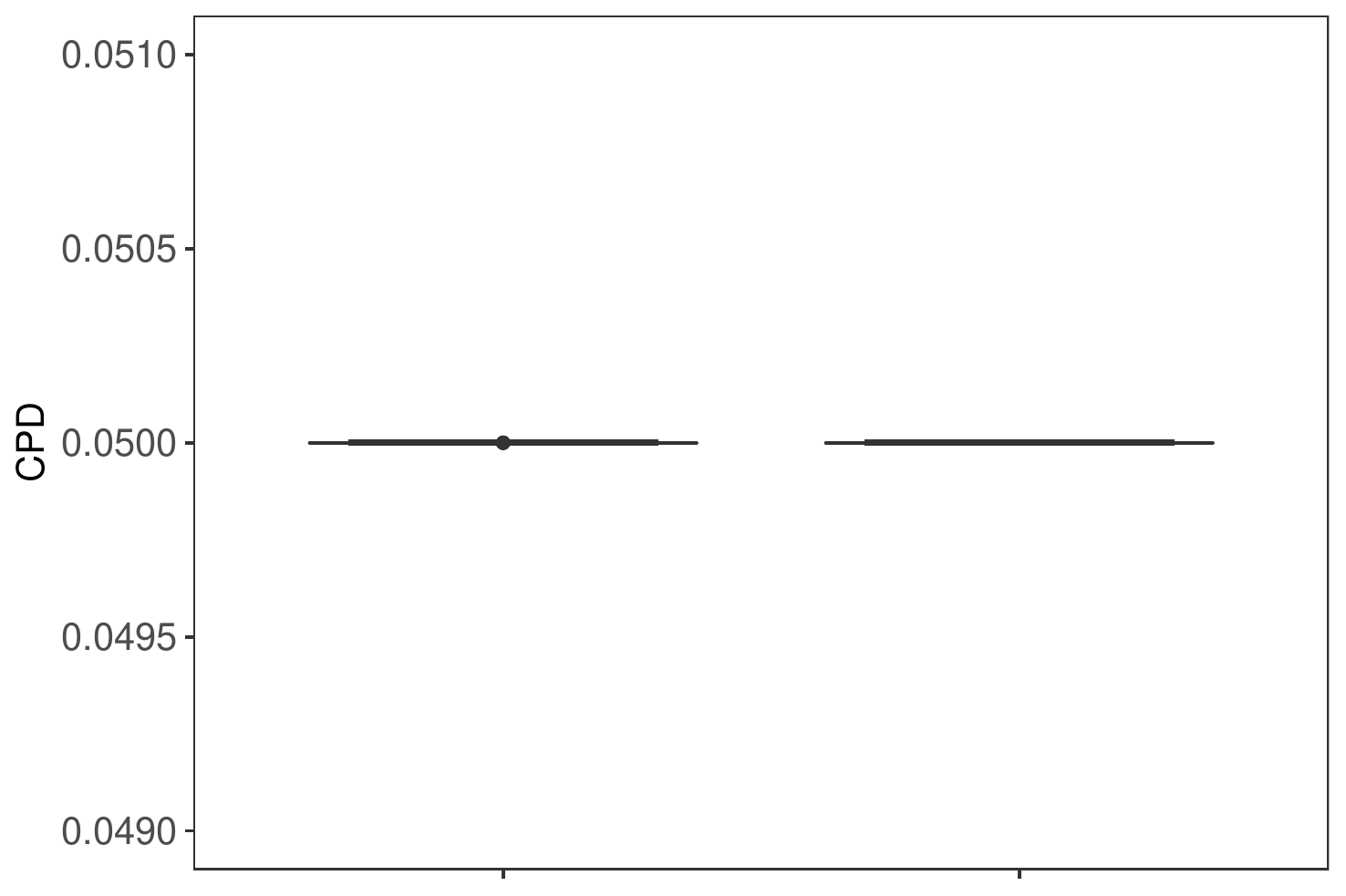}
\qquad
  \includegraphics[width=7.6cm]{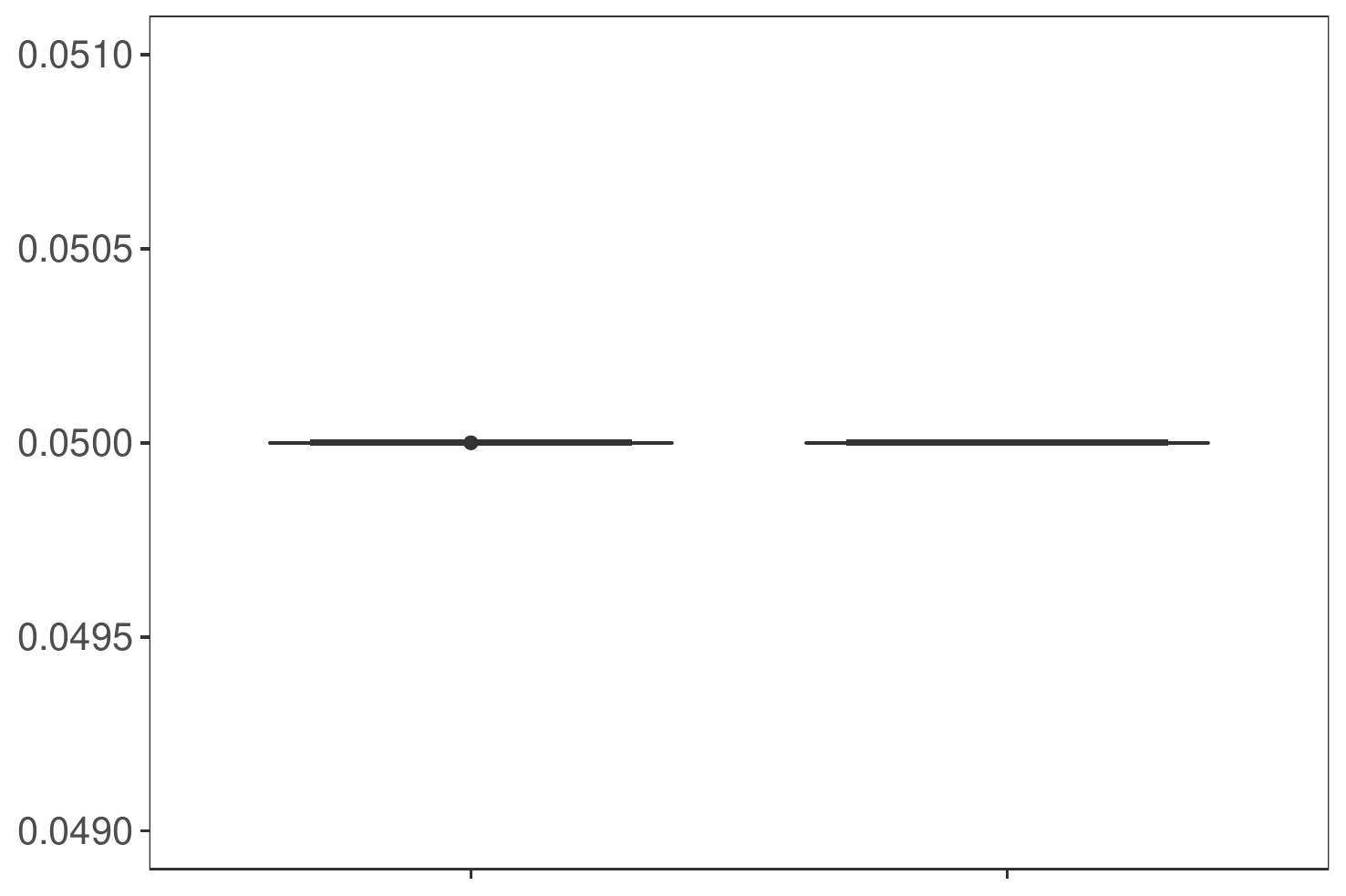}
  \\
  \includegraphics[width=7.6cm]{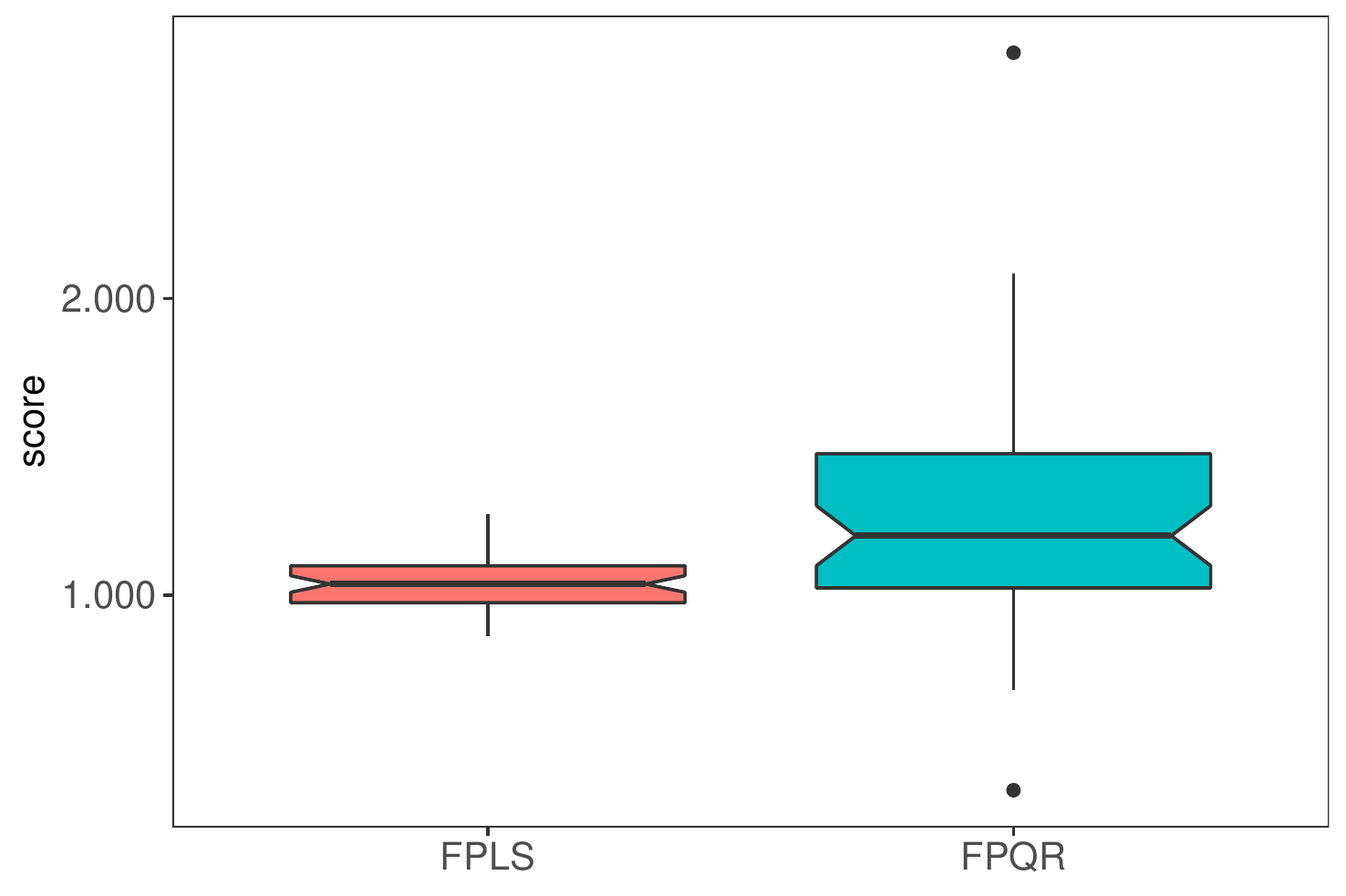}
\qquad
  \includegraphics[width=7.6cm]{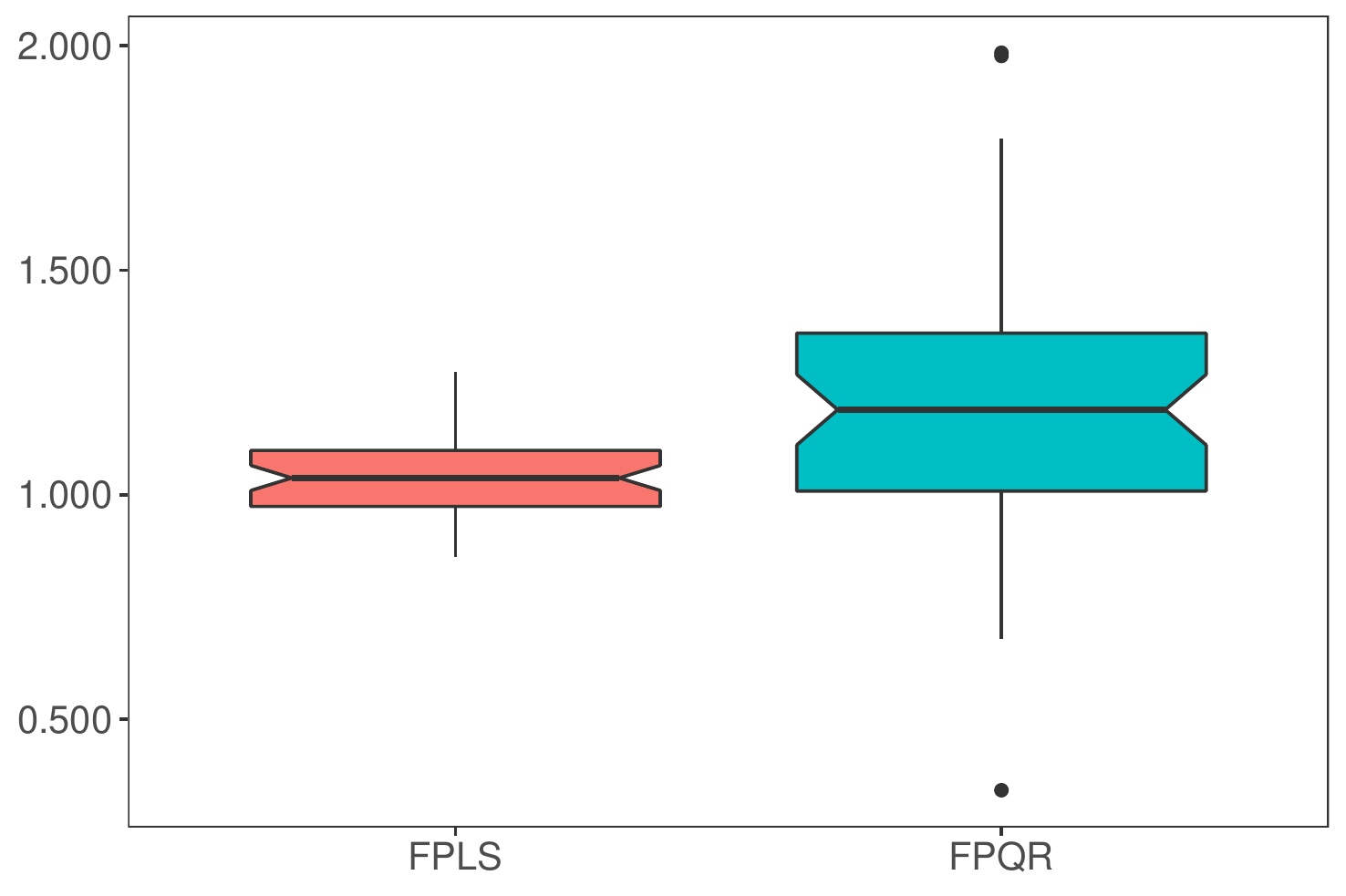}
  \caption{Predictive model performance: MSPE (first row), CPD (second row), and score values (third row) computed under full (first column) and selected (second column) models of the FPLS and FPQR methods for the air quality dataset. The error values for the proposed FPQR are computed when $\tau = 0.5$ levels.}\label{fig:Fig_4}
\end{figure}

The computed MSPE, CPD, and score values obtained from 1000 replications are given in Figure~\ref{fig:Fig_4}. The results demonstrate that both the FPLS and FPQR methods produce smaller error values under \textit{selected model} compared with those calculated under \textit{full model}. This result indicates that the variable selection procedure discussed in Section~\ref{sec:vs} performs well in selecting the significant functional predictors. From Figure~\ref{fig:Fig_4}, the proposed FPQR method with $\tau = 0.5$ produces better performance than the FPLS under the selected model. This result is because the proposed method with quantile parameter $\tau = 0.5$ produces robust predictions for the NO$_2$ curves by reducing the effects of outlying functions. Figure~\ref{fig:Fig_4} also demonstrates that the proposed method produces similar CPD values with slightly larger score values than those of FPLS. 

In addition, a model is constructed by both methods using all 355 curves of all the functional variables to determine the significant variables. Using the fitted models, we calculate the mean squared error (MSE), $\text{MSE} = \frac{1}{355} \sum_{i=1}^{355} \left \Vert \Y_i(t) - \widehat{\Y}_{i}(t) \right \Vert^2_{\mathcal{L}_2}$ where $ \widehat{\Y}_{i}(t)$ denotes the $i\textsuperscript{th}$ fitted NO$_2$ curve, to compare the predictive performance of the FPLS and FPQR. For the FPLS, CO, NMCH, and temperature are selected as significant functional predictors into the final model, and in this case the FPLS produces $\text{MSE} = 0.0050$. On the other hand, all the predictors except C$_6$H$_6$ are selected as significant by the FPQR method (with $\tau = 0.5$), and in this case the calculated MSE is obtained as $\text{MSE} = 0.0026$.

\begin{figure}[!htb]
  \centering
  \includegraphics[width=8cm]{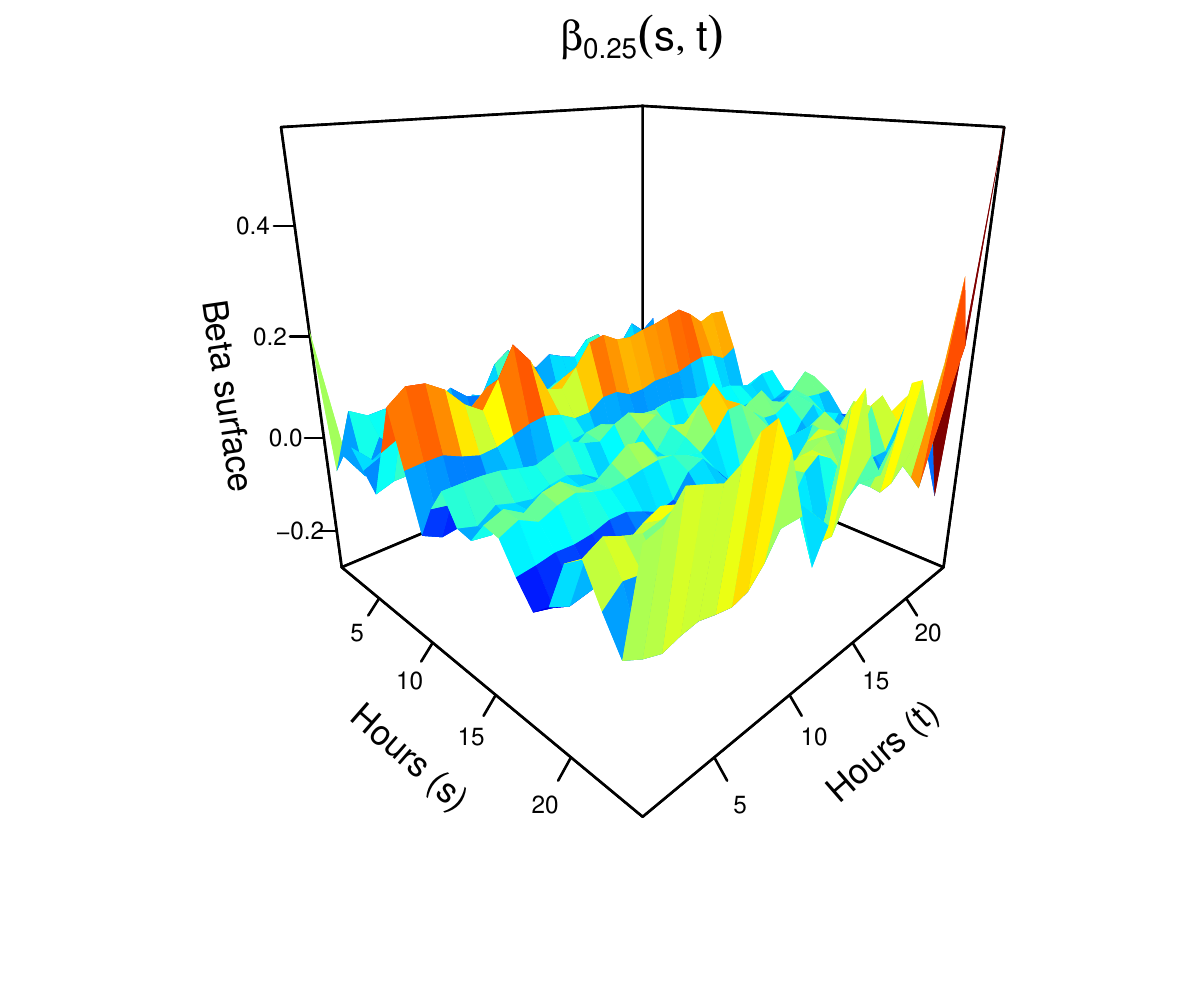}
\quad
  \includegraphics[width=8cm]{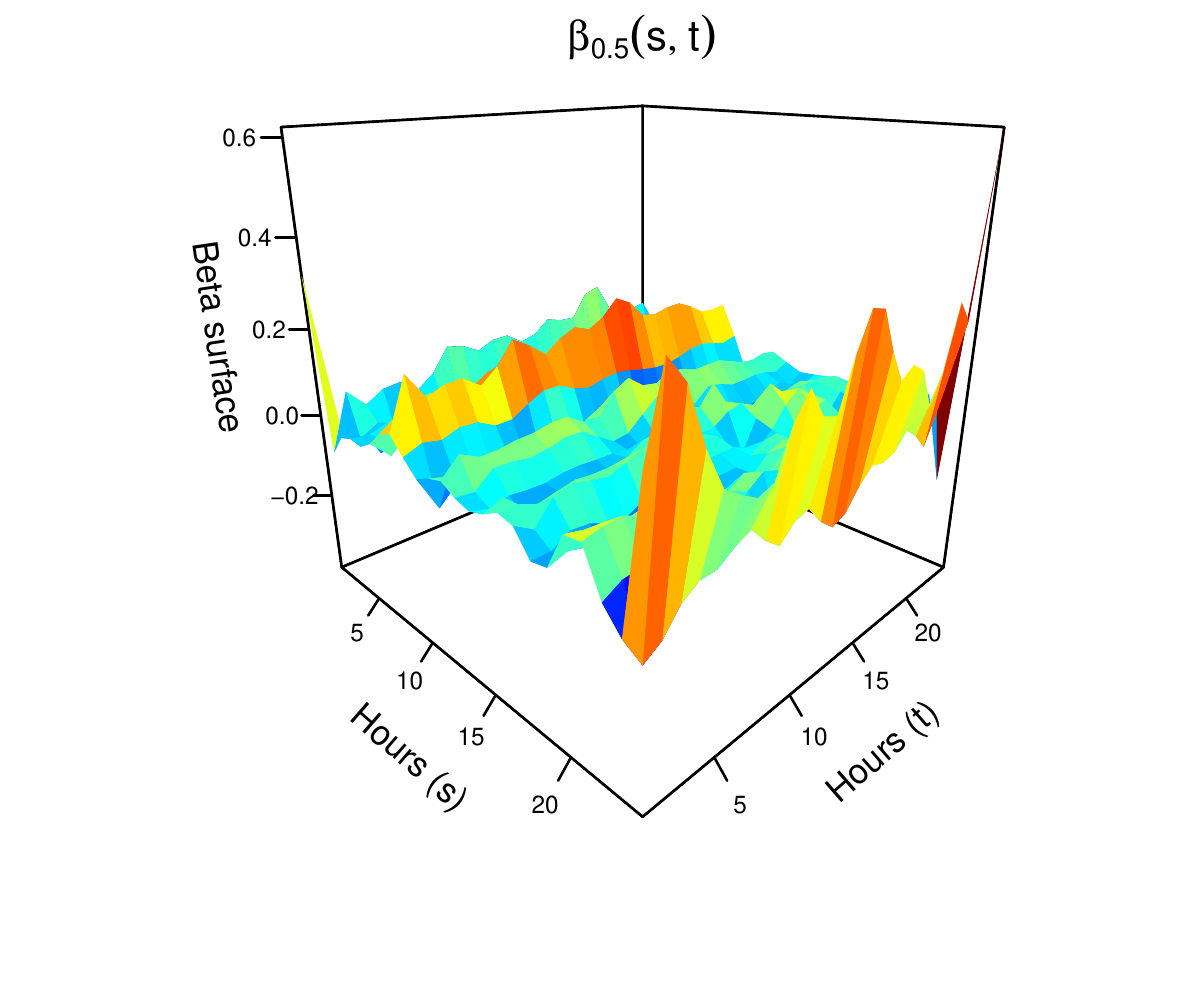}
  \\
  \includegraphics[width=8cm]{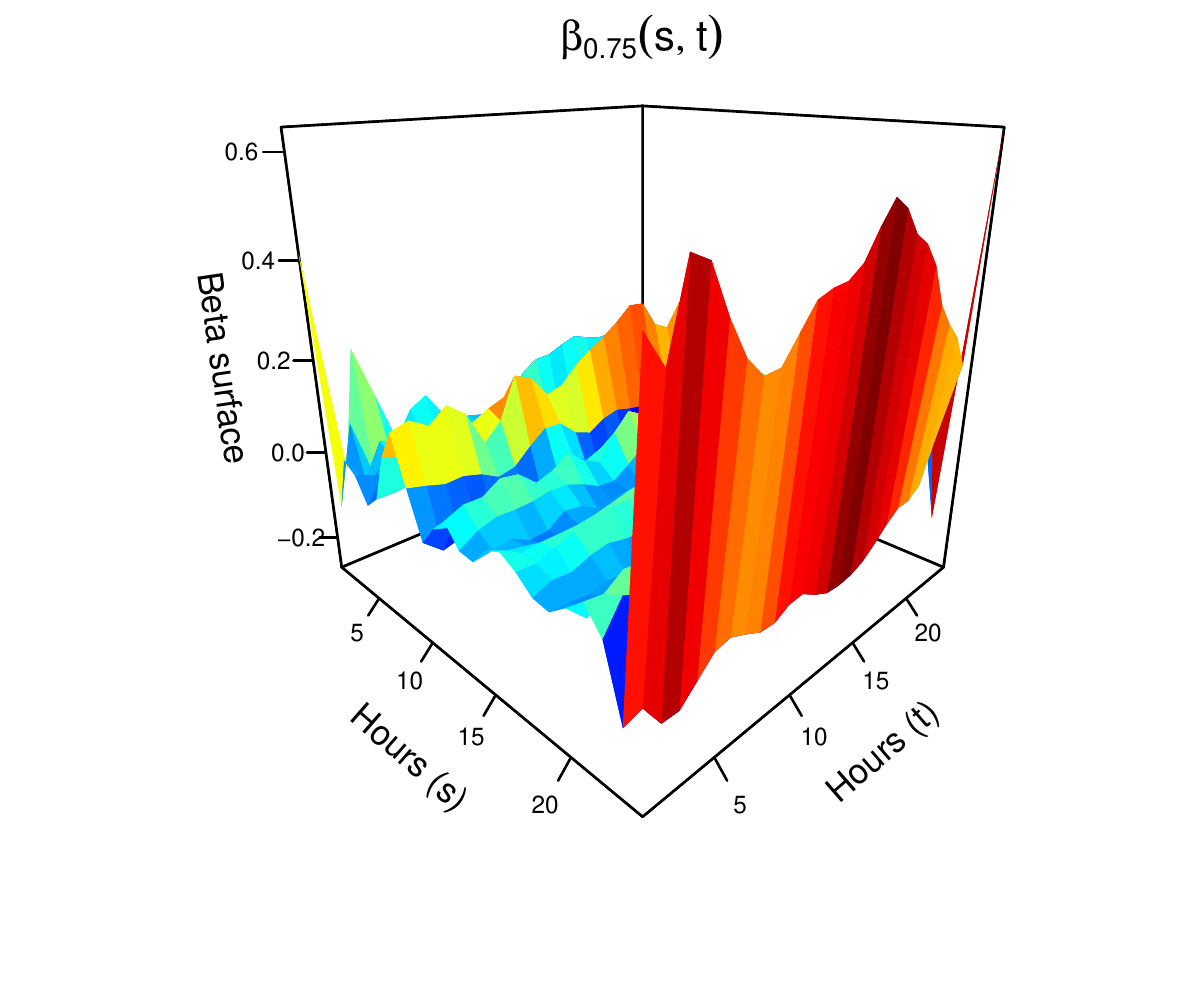}
\quad
  \includegraphics[width=8cm]{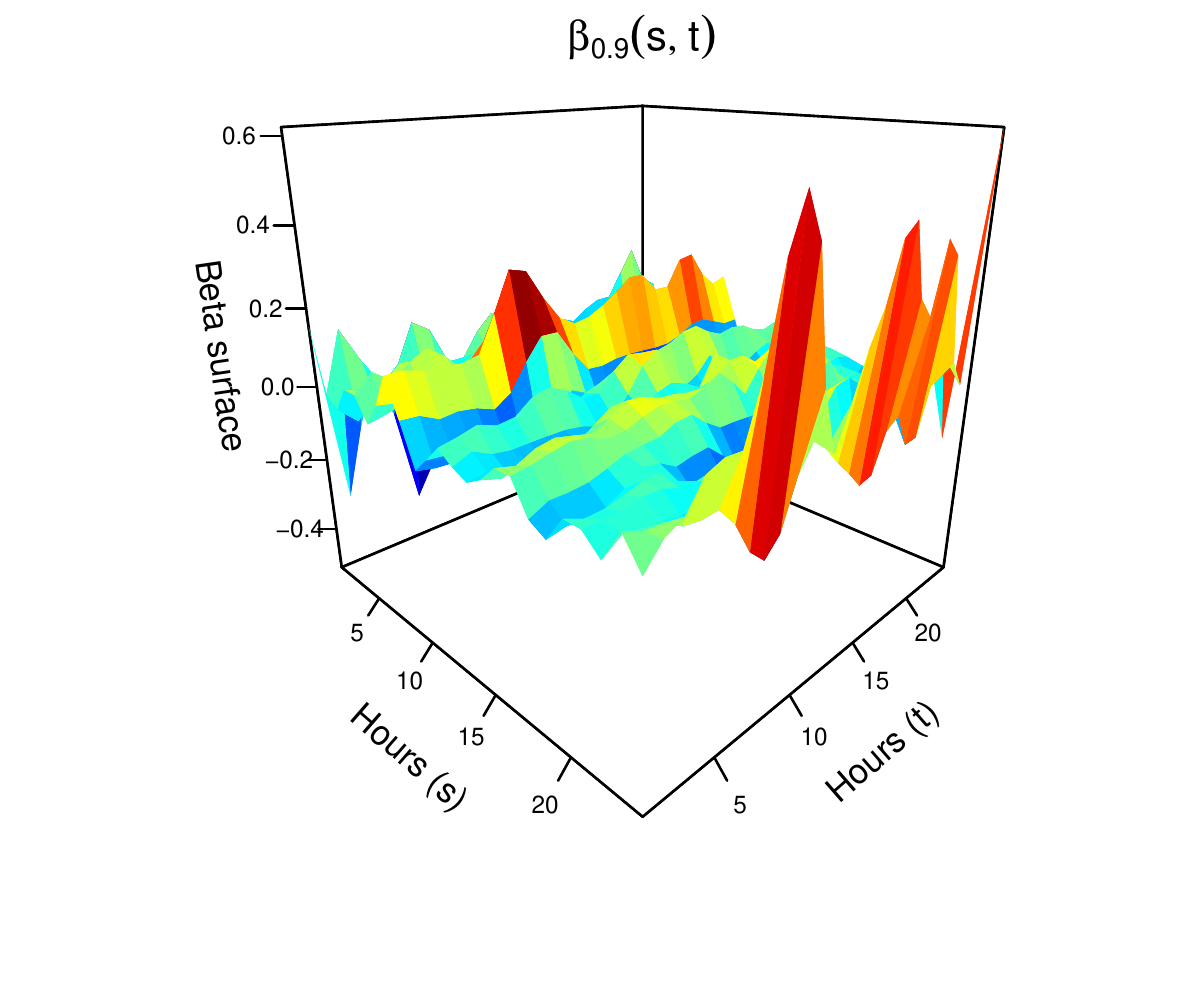}
  \caption{Surface plots of the estimated regression coefficient functions for NMCH when $\tau = [0.25, 0.50, 0.75, 0.90]$.}\label{fig:Fig_5}
\end{figure}

In Figure~\ref{fig:Fig_5}, we present the surface plots of the estimated regression coefficient functions for NMCH (as an example) computed for four $\tau$ levels, $\tau = [0.25, 0.5, 0.75, 0.90]$, to present the effect of NMCH on the different concentration levels of NO$_2$. From Figure~\ref{fig:Fig_5},  it is obvious that the effect of NMCH on NO$_2$ is more significant at upper quantiles, i.e., $\tau = [0.75, 0.90]$, than lower quantiles. In addition, compared with other quantile levels, the effect of NMCH on NO$_2$ is more significant when $\tau = 0.75$.

\section{Conclusion} \label{sec:conc}

A function-on-function linear QR model has been proposed to characterize the functional response's entire conditional distribution for a given set of functional predictors. A FPQR approach is proposed by extending the traditional PQR idea to functional data. The FPQR is an iterative procedure, and in each iteration, uses a partial quantile covariance to extract FPQR basis functions to compute components and estimate the final model. We use the $B$-spline basis expansion method to overcome the ill-posed problem from the functional random variables' infinite-dimensional nature. The proposed FPQR constructed using the functional random variables is approximated via the multivariate PQR constructed using the basis expansion coefficients. A Bayesian information criterion is used to determine the optimum number of retained FPQR components. A forward variable selection is used to select only the significant functional predictors in the final model.

The predictive performance of the proposed method is evaluated via several Monte Carlo experiments and empirical data analysis, and the performance of the FPQR are compared with those of FPLS. Our results have demonstrated that the proposed method produces improved accuracy than the FPLS when the data include outliers or errors that follow a non-Gaussian heavy-tailed distribution. Under the Gaussian case, it delivers competitive performance with the FPLS. 

We present some ideas in which the proposed method can be further extended:
\begin{inparaenum}
\item[1)] We consider only the $B$-spline basis expansion method to construct the discretely observed data's functional forms. However, the predictive performance of the proposed method may depend on the selected basis expansion method. Recently, \cite{Wang2019} and \cite{Yu2019} proposed scalar-on-function linear quantile regression models based on a set of wavelet bases and \cite{Sang2020} proposed a functional single-index quantile regression model based on a set of Fourier bases. Similar to these cases, the performance of the proposed FPQR method on the function-on-function linear quantile regression can be explored using other basis expansion methods, such as Fourier, wavelet, radial, and Bernstein polynomial bases.
\item[2)] We consider only the main effects of the functional predictors. However, recent studies have shown that functional regression models, including a quadratic term and interaction effects, perform better than standard functional regression models in the presence of interaction \citep{LuoQi, Matsui2020, SunWang, BeyaztasShang2021}. The functional predictors' quadratic or interaction effects can also be used in the proposed method to characterize the functional response's conditional distribution.
\end{inparaenum}



\bibliographystyle{agsm}
\bibliography{pfqr}

\end{document}